\documentclass[article,nojss]{jss}
\usepackage{amsmath}
\usepackage[plain]{algorithm}
\usepackage{algorithmic}
\usepackage{bm}
\usepackage{xspace}

\def\vsmc{\pkg{vSMC}\xspace}

\def\caocl{AMD \pkg{APP} \proglang{OpenCL}\xspace\citep{aocl}\xspace}
\def\cboost{\pkg{Boost}\xspace\citep{boost}\xspace}
\def\cclang{\pkg{clang}\xspace\citep{clang}\xspace}
\def\ccmake{\pkg{CMake}\xspace\citep{cmake}\xspace}
\def\cdoxygen{\pkg{Doxygen}\xspace\citep{doxygen}\xspace}
\def\cgcc{\pkg{GCC}\xspace\citep{gcc}\xspace}
\def\cgit{\pkg{Git}\xspace\citep{git}\xspace}
\def\cicpc{Intel \proglang{C++} Complier\xspace\citep{icpc}\xspace}
\def\ciocl{Intel \proglang{OpenCL}\xspace\citep{iocl}\xspace}
\def\clibcpp{\pkg{libc++}\xspace\citep{libcpp}\xspace}
\def\cmsvc{Microsoft \proglang{Visual C++}\xspace\citep{msvc}\xspace}
\def\crandom{\pkg{Random123}\xspace\citep{random}\xspace}
\def\crlang{\pkg{R}\xspace\citep{rlang}\xspace}
\def\csmctc{\pkg{SMCTC}\xspace\citep{smctc}\xspace}
\def\ctbb{\pkg{Intel TBB}\xspace\citep{tbb}\xspace}

\def\cbiips{\proglang{BiiPS}\xspace\citep{biips}\xspace}
\def\cbugs{\proglang{BUGS}\xspace\citep{bugs}\xspace}
\def\ccilk{\proglang{Cilk Plus}\xspace\citep{cilk}\xspace}
\def\clibbi{\proglang{LibBi}\xspace\citep{Murray2013bi}\xspace}
\def\cmpi{\proglang{MPI}\xspace\citep{mpi}\xspace}
\def\copencl{\proglang{OpenCL}\xspace\citep{opencl}\xspace}
\def\copenmp{\proglang{OpenMP}\xspace\citep{openmp}\xspace}

\def\aocl{AMD \pkg{APP} \proglang{OpenCL}\xspace}
\def\boost{\pkg{Boost}\xspace}
\def\clang{\pkg{clang}\xspace}
\def\cmake{\pkg{CMake}\xspace}

\def\gcc{\pkg{GCC}\xspace}

\def\icpc{Intel \proglang{C++} Complier\xspace}
\def\iocl{Intel \proglang{OpenCL}\xspace}
\def\libcpp{\pkg{libc++}\xspace}
\def\msvc{Microsoft \proglang{Visual C++}\xspace}

\def\rlang{\pkg{R}\xspace}
\def\smctc{\pkg{SMCTC}\xspace}
\def\tbb{\pkg{Intel TBB}\xspace}

\def\bugs{\proglang{BUGS}\xspace}
\def\cilk{\proglang{Cilk Plus}\xspace}

\def\mpi{\proglang{MPI}\xspace}
\def\opencl{\proglang{OpenCL}\xspace}
\def\openmp{\proglang{OpenMP}\xspace}

\def\cppne{\proglang{C++98}\xspace}
\def\cppoo{\proglang{C++11}\xspace}
\def\cpp{\proglang{C++}\xspace}
\def\html{\proglang{HTML}\xspace}

\def\ais{AIS\xspace}
\def\cpu{CPU\xspace}
\def\gmm{GMM\xspace}
\def\gpgpu{GPGPU\xspace}
\def\gpu{GPU\xspace}

\def\io{IO\xspace}
\def\mcmc{MCMC\xspace}
\def\rng{RNG\xspace}
\def\simd{SIMD\xspace}
\def\sis{SIS\xspace}
\def\smc{SMC\xspace}
\def\smp{SMP\xspace}
\def\stl{STL\xspace}

\DeclareMathOperator{\diff}{d}
\def\Exp{E}
\def\Real{R}
\def\calM{\mathcal{M}}
\def\data{\bm{y}}
\def\intd{\,\diff}
\def\rdir{\mathcal{D}}
\def\rgamma{\mathcal{G}}
\def\rnorm{\mathcal{N}}
\def\rnd#1#2{\ensuremath{\frac{\diff#1}{\diff#2}}\xspace}
\def\ess{\ifmmode\text{ESS}\else{ESS}\xspace\fi}

\def\xpos{x_{\mathrm{pos}}}
\def\ypos{y_{\mathrm{pos}}}
\def\xvel{x_{\mathrm{vel}}}
\def\yvel{y_{\mathrm{vel}}}
\def\xobs{x_{\mathrm{obs}}}
\def\yobs{y_{\mathrm{obs}}}

\def\STATESKIP{\hskip.68cm}

\title{vSMC: Parallel Sequential Monte Carlo in \cpp}
\author{Yan Zhou\\University of Warwick}

\Plaintitle{vSMC: Parallel Sequential Monte Carlo in C++}
\Plainauthor{Zhou, Yan}

\Abstract
{
  Sequential Monte Carlo is a family of algorithms for sampling from a
  sequence of distributions. Some of these algorithms, such as particle
  filters, are widely used in the physics and signal processing researches.
  More recent developments have established their application in more general
  inference problems such as Bayesian modeling.

  These algorithms have attracted considerable attentions in recent years as
  they admit natural and scalable parallelizations. However, these algorithms
  are perceived to be difficult to implement. In addition, parallel
  programming is often unfamiliar to many researchers though conceptually
  appealing, especially for sequential Monte Carlo related fields.

  A \cpp template library is presented for the purpose of implementing general
  sequential Monte Carlo algorithms on parallel hardware. Two examples are
  presented: a simple particle filter and a classic Bayesian modeling problem.
}

\Plainkeywords{Monte Carlo, particle filter, C++ template generic programming}

\Address
{
  Yan Zhou\\
  Department of Statistics\\
  University of Warwick\\
  Coventry, CV4 7AL, United Kingdom\\
  E-mail: \email{Yan.Zhou@warwick.ac.uk}\\
  URL: \url{http://zhouyan.github.com}
}

\begin{document}

\section{Introduction}
\label{sec:Introduction}

Sequential Monte Carlo (\smc) methods are a class of sampling algorithms that
combine importance sampling and resampling. They have been primarily used as
``particle filters'' to solve optimal filtering problems; see, for example,
\citet{Cappe:2007hz} and \citet{Doucet:2011us} for recent reviews. They are
also used in a static setting where a target distribution is of interest, for
example, for the purpose of Bayesian modeling. This was proposed by
\citet{DelMoral:2006hc} and developed by \citet{Peters:2005wh} and
\citet{DelMoral:2006wv}. This framework involves the construction of a
sequence of artificial distributions on spaces of increasing dimensions which
admit the distributions of interest as particular marginals.

\smc algorithms are perceived as being difficult to implement while general
tools were not available until the development by \citet{smctc}, which
provided a general framework for implementing \smc algorithms. \smc algorithms
admit natural and scalable parallelization. However, there are only parallel
implementations of \smc algorithms for many problem specific applications,
usually associated with specific \smc related researches. \citet{Lee:2010fm}
studied the parallelization of \smc algorithms on \gpu{}s with some
generality. There are few general tools to implement \smc algorithms on
parallel hardware though multicore \cpu{}s are very common today and computing
on specialized hardware such as \gpu{}s are more and more popular.

The purpose of the current work is to provide a general framework for
implementing \smc algorithms on both sequential and parallel hardware. There
are two main goals of the presented framework. The first is reusability. It
will be demonstrated that the same implementation source code can be used to
build a serialized sampler, or using different programming models (for
example, \copenmp and \ctbb) to build parallelized samplers for multicore
\cpu{}s. They can be scaled for clusters using \cmpi with few modifications.
And with a little effort they can also be used to build parallelized samplers
on specialized massive parallel hardware such as \gpu{}s using \copencl. The
second is extensibility. It is possible to write a backend for \vsmc to use
new parallel programming models while reusing existing implementations. It is
also possible to enhance the library to improve performance for specific
applications. Almost all components of the library can be reimplemented by
users and thus if the default implementation is not suitable for a specific
application, they can be replaced while being integrated with other components
seamlessly.

\section{Sequential Monte Carlo}
\label{sec:Sequential Monte Carlo}

\subsection{Sequential importance sampling and resampling}
\label{sub:Sequential importance sampling and resampling}

Importance sampling is a technique which allows the calculation of the
expectation of a function $\varphi$ with respect to a distribution $\pi$ using
samples from some other distribution $\eta$ with respect to which $\pi$ is
absolutely continuous, based on the identity,
\begin{equation}
  \Exp_{\pi}[\varphi(X)]
  = \int\varphi(x)\pi(x)\intd x
  = \int\frac{\varphi(x)\pi(x)}{\eta(x)}\eta(x)\intd x
  = \Exp_{\eta}\Bigl[\frac{\varphi(X)\pi(X)}{\eta(X)}\Bigr]
\end{equation}
And thus, let $\{X^{(i)}\}_{i=1}^N$ be samples from $\eta$, then
$\Exp_{\pi}[\varphi(X)]$ can be approximated by
\begin{equation}
  \hat\varphi_1 =
  \frac{1}{N}\sum_{i=1}^N\frac{\varphi(X^{(i)})\pi(X^{(i)})}{\eta(X^{(i)})}
\end{equation}
In practice $\pi$ and $\eta$ are often only known up to some normalizing
constants, which can be estimated using the same samples. Let $w^{(i)} =
\pi(X^{(i)})/\eta(X^{(i)})$, then we have
\begin{equation}
  \hat\varphi_2 =
  \frac{\sum_{i=1}^Nw^{(i)}\varphi(X^{(i)})}{\sum_{i=1}^Nw^{(i)}}
\end{equation}
or
\begin{equation}
  \hat\varphi_3 = \sum_{i=1}^NW^{(i)}\varphi(X^{(i)})
\end{equation}
where $W^{(i)}\propto w^{(i)}$ and are normalized such that
$\sum_{i=1}^NW^{(i)} = 1$.

Sequential importance sampling (\sis) generalizes the importance sampling
technique for a sequence of distributions $\{\pi_t\}_{t\ge0}$ defined on
spaces $\{\prod_{k=0}^tE_k\}_{t\ge0}$. At time $t = 0$, sample
$\{X_0^{(i)}\}_{i=1}^N$ from $\eta_0$ and compute the weights $W_0^{(i)}
\propto \pi_0(X_0^{(i)})/\eta_0(X_0^{(i)})$. At time $t\ge1$, each sample
$X_{0:t-1}^{(i)}$, usually termed \emph{particles} in the literature, is
extended to $X_{0:t}^{(i)}$ by a proposal distribution
$q_t(\cdot|X_{0:t-1}^{(i)})$. And the weights are recalculated by $W_t^{(i)}
\propto \pi_t(X_{0:t}^{(i)})/\eta_t(X_{0:t}^{(i)})$ where
\begin{equation}
  \eta_t(X_{0:t}^{(i)}) =
  \eta_{t-1}(X_{0:t-1}^{(i)})q_t(X_{0:t}^{(i)}|X_{0:t-1}^{(i)})
\end{equation}
and thus
\begin{align}
  W_t^{(i)} \propto \frac{\pi_t(X_{0:t}^{(i)})}{\eta_t(X_{0:t}^{(i)})}
  &= \frac{\pi_t(X_{0:t}^{(i)})\pi_{t-1}(X_{0:t-1}^{(i)})}
  {\eta_{t-1}(X_{0:t-1}^{(i)})q_t(X_{0:t}^{(i)}|X_{0:t-1}^{(i)})
    \pi_{t-1}(X_{0:t-1}^{(i)})} \notag\\
  &= \frac{\pi_t(X_{0:t}^{(i)})}
  {q_t(X_{0:t}^{(i)}|X_{0:t-1}^{(i)})\pi_{t-1}(X_{0:t-1}^{(i)})}W_{t-1}^{(i)}
  \label{eq:si}
\end{align}
and importance sampling estimate of $\Exp_{\pi_t}[\varphi_t(X_{0:t})]$ can be
obtained using $\{W_t^{(i)},X_{0:t}^{(i)}\}_{i=1}^N$.

However this approach fails as $t$ becomes large. The weights tend to become
concentrated on a few particles as the discrepancy between $\eta_t$ and
$\pi_t$ becomes larger. Resampling techniques are applied such that, a new
particle system $\{\bar{W}_t^{(i)},\bar{X}_{0:t}^{(i)}\}_{i=1}^M$ is obtained
with the property,
\begin{equation}
  \Exp\Bigl[\sum_{i=1}^M\bar{W}_t^{(i)}\varphi_t(\bar{X}_{0:t}^{(i)})\Bigr] =
  \Exp\Bigl[\sum_{i=1}^NW_t^{(i)}\varphi_t(X_{0:t}^{(i)})\Bigr]
  \label{eq:resample}
\end{equation}
In practice, the resampling algorithm is usually chosen such that $M = N$ and
$\bar{W}^{(i)} = 1/N$ for $i=1,\dots,N$. Resampling can be performed at each time
$t$ or adaptively based on some criteria of the discrepancy. One popular
quantity used to monitor the discrepancy is \emph{effective sample size}
(\ess), introduced by \citet{Liu:1998iu}, defined as
\begin{equation}
  \ess_t = \frac{1}{\sum_{i=1}^N (W_t^{(i)})^2}
\end{equation}
where $\{W_t^{(i)}\}_{i=1}^N$ are the normalized weights. And resampling can
be performed when $\ess\le \alpha N$ where $\alpha\in[0,1]$.

The common practice of resampling is to replicate particles with large weights
and discard those with small weights. In other words, instead of generating a
random sample $\{\bar{X}_{0:t}^{(i)}\}_{i=1}^N$ directly, a random sample of
integers $\{R^{(i)}\}_{i=1}^N$ is generated, such that $R^{(i)} \ge 0$ for $i
= 1,\dots,N$ and $\sum_{i=1}^N R^{(i)} = N$. And each particle value
$X_{0:t}^{(i)}$ is replicated for $R^{(i)}$ times in the new particle system.
The distribution of $\{R^{(i)}\}_{i=1}^N$ shall fulfill the requirement of
Equation~\ref{eq:resample}. One such distribution is a multinomial
distribution of size $N$ and weights $(W_t^{(i)},\dots,W_t^{(N)})$. See
\citet{Douc:2005wa} for some commonly used resampling algorithms.

\subsection[SMC samplers]{\protect\smc samplers}
\label{sub:SMC Samplers}

\smc samplers allow us to obtain, iteratively, collections of weighted samples
from a sequence of distributions $\{\pi_t\}_{t\ge0}$ over essentially any
random variables on some spaces $\{E_t\}_{t\ge0}$, by constructing a sequence
of auxiliary distributions $\{\tilde\pi_t\}_{t\ge0}$ on spaces of increasing
dimensions, $\tilde\pi_t(x_{0:t})=\pi_t (x_t) \prod_{s=0}^{t-1}
L_s(x_{s+1},x_s)$, where the sequence of Markov kernels $\{L_s\}_{s=0}^{t-1}$,
termed backward kernels, is formally arbitrary but critically influences the
estimator variance. See \citet{DelMoral:2006hc} for further details and
guidance on the selection of these kernels.

Standard sequential importance sampling and resampling algorithms can then be
applied to the sequence of synthetic distributions, $\{\tilde\pi_t\}_{t\ge0}$.
At time $t - 1$, assume that a set of weighted particles
$\{W_{t-1}^{(i)},X_{0:t-1}^{(i)}\}_{i=1}^N$ approximating $\tilde\pi_{t-1}$ is
available, then at time $t$, the path of each particle is extended with a
Markov kernel say, $K_t(x_{t-1}, x_t)$ and the set of particles
$\{X_{0:t}^{(i)}\}_{i=1}^N$ reach the distribution $\eta_t(X_{0:t}^{(i)}) =
\eta_0(X_0^{(i)})\prod_{k=1}^tK_t(X_{t-1}^{(i)}, X_t^{(i)})$, where $\eta_0$
is the initial distribution of the particles. To correct the discrepancy
between $\eta_t$ and $\tilde\pi_t$, Equation~\ref{eq:si} is applied and in
this case,
\begin{equation}
  W_t^{(i)} \propto \frac{\tilde\pi_t(X_{0:t}^{(i)})}{\eta_t(X_{0:t}^{(i)})}
  = \frac{\pi_t(X_t^{(i)})\prod_{s=0}^{t-1}L_s(X_{s+1}^{(i)}, X_s^{(i)})}
  {\eta_0(X_0^{(i)})\prod_{k=1}^tK_t(X_{t-1}^{(i)},X_t^{(i)})}
  \propto \tilde{w}_t(X_{t-1}^{(i)}, X_t^{(i)})W_{t-1}^{(i)}
\end{equation}
where $\tilde{w}_t$, termed the \emph{incremental weights}, are calculated as,
\begin{equation}
  \tilde{w}_t(X_{t-1}^{(i)},X_t^{(i)}) =
  \frac{\pi_t(X_t^{(i)})L_{t-1}(X_t^{(i)}, X_{t-1}^{(i)})}
  {\pi_{t-1}(X_{t-1}^{(i)})K_t(X_{t-1}^{(i)}, X_t^{(i)})}
\end{equation}
If $\pi_t$ is only known up to a normalizing constant, say $\pi_t(x_t) =
\gamma_t(x_t)/Z_t$, then we can use the \emph{unnormalized} incremental
weights
\begin{equation}
  w_t(X_{t-1}^{(i)},X_t^{(i)}) =
  \frac{\gamma_t(X_t^{(i)})L_{t-1}(X_t^{(i)}, X_{t-1}^{(i)})}
  {\gamma_{t-1}(X_{t-1}^{(i)})K_t(X_{t-1}^{(i)}, X_t^{(i)})}
\end{equation}
for importance sampling. Further, with the previously \emph{normalized}
weights $\{W_{t-1}^{(i)}\}_{i=1}^N$, we can estimate the ratio of normalizing
constant $Z_t/Z_{t-1}$ by
\begin{equation}
  \frac{\hat{Z}_t}{Z_{t-1}} =
  \sum_{i=1}^N W_{t-1}^{(i)}w_t(X_{t-1}^{(i)},X_t^{(i)})
\end{equation}
Sequentially, the normalizing constant between initial distribution $\pi_0$
and some target $\pi_T$, $T\ge1$ can be estimated. See \citet{DelMoral:2006hc}
for details on calculating the incremental weights. In practice, when $K_t$ is
invariant to $\pi_t$, and an approximated suboptimal backward kernel
\begin{equation}
  L_{t-1}(x_t, x_{t-1}) = \frac{\pi(x_{t-1})K_t(x_{t-1}, x_t)}{\pi_t(x_t)}
\end{equation}
is used, the unnormalized incremental weights will be
\begin{equation}
  w_t(X_{t-1}^{(i)},X_t^{(i)}) =
  \frac{\gamma_t(X_{t-1}^{(i)})}{\gamma_{t-1}(X_{t-1}^{(i)})}.
  \label{eq:inc_weight_mcmc}
\end{equation}

\subsection{Other sequential Monte Carlo algorithms}
\label{sub:Other sequential Monte Carlo algorithms}

Some other commonly used sequential Monte Carlo algorithms can be viewed as
special cases of algorithms introduced above. The annealed importance sampling
(\ais; \citet{Neal:2001we}) can be viewed as \smc samplers without resampling.

Particle filters as seen in the physics and signal processing literature, can
also be interpreted as the sequential importance sampling and resampling
algorithms. See \citet{Doucet:2011us} for a review of this topic. A simple
particle filter example is used in Section~\ref{sub:A simple particle filter}
to demonstrate basic features of the \vsmc library.

\section{Using the vSMC library}
\label{sec:Using the vSMC library}

\subsection{Overview}

The \vsmc library makes use of \cpp's template generic programming to
implement general \smc algorithms. This library is formed by a few major
modules listed below. Some features not included in these modules are
introduced later in context.
\begin{description}
  \item[Core] The highest level of abstraction of \smc samplers. Users
    interact with classes defined within this module to create and manipulate
    general \smc samplers. Classes in this module include \code{Sampler},
    \code{Particle} and others. These classes use user defined callback
    functions or callable objects, such as functors, to perform problem
    specific operations, such as updating particle values and weights. This
    module is documented in Section~\ref{sub:Core module}.
  \item[Symmetric Multiprocessing (\smp)] This is the form of computing most
    people use everyday, including multiprocessor workstations, multicore
    desktops and laptops. Classes within this module make it possible to write
    generic operations which manipulate a single particle that can be applied
    either sequentially or in parallel through various parallel programming
    models. A method defined through classes of this module can be used by
    \code{Sampler} as callback objects. This module is documented in
    Section~\ref{sub:SMP module}.
  \item[Message Passing Interface] \mpi is the \emph{de facto} standard
    for parallel programming on distributed memory architectures. This module
    enables users to adapt implementations of algorithms written for the \smp
    module such that the same sampler can be parallelized using \mpi. In
    addition, when used with the \smp module, it allows easy implementation of
    hybrid parallelization such as \mpi/\openmp. In Section~\ref{sub:Bayesian
      Modeling of Gaussian mixture model}, an example is shown how to extend
    existing \smp parallelized samplers using this module.
  \item[OpenCL] This module is similar to the two above except it eases the
    parallelization through \opencl, such as for the purpose of General
    Purpose \gpu Programming (\gpgpu). \opencl is a framework for writing
    programs that can be execute across heterogeneous platforms. \opencl
    programs can run on either \cpu{}s or \gpu{}s. It is beyond the scope of
    this paper to give a proper introduction to \gpgpu, \opencl and their use
    in \vsmc. However, later we will demonstrate the relative performance of
    this programming model on both \cpu{}s and \gpu{}s in
    Section~\ref{sub:Bayesian Modeling of Gaussian mixture model}.
\end{description}

\subsection{Obtaining and installation}

\vsmc is a header only library. There is practically no installation step. The
library can be downloaded from \url{http://zhouyan.github.io/vSMC/vSMC.zip}.
After downloading and unpacking, one can start using \vsmc by ensuring that
the compiler can find the headers inside the \code{include} directory. To
permanently install the library in a system directory, one can simply copy the
contents of the \code{include} directory to an appropriate place.

Alternatively, one can use the \ccmake (2.8 or later) configuration script and
obtain the source by \cgit. On a Unix-like system (such as Mac OS X, BSD,
Linux and others),
\begin{Code}
git clone git://github.com/zhouyan/vSMC.git
cd vSMC
git submodule init
git submodule update
mkdir build
cd build
cmake .. -DCMAKE_BUILD_TYPE=Release
make install
\end{Code}
For Unix-like systems, there is also a shell script \code{build.sh} that
builds all examples in this paper and produces the results and benchmarks as
in Section~\ref{sec:Example}. See documentations in that script for details of
how to change settings for the users' platform.

\subsection{Documentation}

To build the reference manual, one need \cdoxygen, version~1.8.3 or later.
Continuing the last step (still in the \code{build} directory), invoking
\code{make docs} will create a \code{doc} directory inside \code{build}, which
contains the \html references. Alternatively the reference manual can also be
found on \url{http://zhouyan.github.io/vSMC/doc/html/index.html}. It is beyond
the scope of this paper to document every feature of the \vsmc library. In
many places we will refer to this reference manual for further information.

\subsection{Third-party dependencies}

\vsmc uses \crandom counter-based \rng for random number generating. For an
\smc sampler with $N$ particles, \vsmc constructs $N$ (statistically)
independent \rng streams. It is possible to use millions of such streams
without a huge memory footprint or other performance penalties. Since each
particle has its own independent \rng stream, it frees users from many
thread-safety and statistical independence considerations. It is highly
recommended that the users install this library. Within \vsmc, these \rng
streams are wrapped under \cppoo \rng engines, and can be replaced by other
compatible \rng engines seamlessly. Users only need to be familiar with
classes defined in \cppoo{} \code{<random>} or their \cboost equivalents to
use these \rng streams. See the documentation of the corresponding libraries
for details, as well as examples in Section~\ref{sec:Example}.

The other third-party dependency is the \boost library. Version 1.49 or later
is required. However, this is actually optional provided that proper \cppoo
features are available in the standard library, for example using \cclang with
\clibcpp. The \cppoo headers of concern are \code{<functional>} and
\code{<random>}. To instruct \vsmc to use the standard library headers instead
of falling back to the \boost library, one needs to define the configuration
macros before including any \vsmc headers. For example,
\begin{Code}
clang++ -std=c++11 -stdlib=libc++     \
    -DVSMC_HAS_CXX11LIB_FUNCTIONAL=1  \
    -DVSMC_HAS_CXX11LIB_RANDOM=1      \
    -o prog prog.cpp
\end{Code}
tells the library to use \cppoo{} \code{<functional>} and \code{<random>}.
The availability of these headers are also checked by the \cmake configuration
script.

\subsection{Compiler support}

\vsmc has been tested with recent versions of \clang, \cgcc, \cicpc and
\cmsvc. \vsmc can optionally use some \cppoo features to improve performance
and usability. In particular, as mentioned before, \vsmc can use \cppoo
standard libraries instead of the \boost library. At the time of writing,
\clang with \libcpp has the most comprehensive support of \cppoo with respect
to standard compliant and feature completion. \gcc 4.8 , \msvc 2012 and \icpc
2013 also have very good \cppoo support. Note that, provided the \boost
library is available, all \cppoo language and library features are optional.
\vsmc can be used with any \cppne conforming compilers.

\section{The vSMC library}
\label{sec:The vSMC library}

\subsection{Core module}
\label{sub:Core module}

The core module abstracts general \smc samplers. \smc samplers can be viewed
as formed by a few concepts regardless of the specific problems. The following
is a list of the most commonly seen components of \smc samplers and their
corresponding \vsmc abstractions.
\begin{itemize}
  \item A collection of all particle state values, namely
    $\{X_t^{(i)}\}_{i=1}^N$. In \vsmc, users need to define a class, say
    \code{T}, to abstract this concept. We refer to this as the \emph{value
      collection}. We will slightly abuse the generic name \code{T} in this
    paper. Whenever a template parameter is mentioned with the name \code{T},
    it always refers to such a value collection type unless stated otherwise.
  \item A collection of all particle state values and their associated
    weights. This is abstracted by a \code{Particle<T>} object. We refer to
    this as the \emph{particle collection}. A \code{Particle<T>} object has
    three primary sub-objects. One is the above type \code{T} value collection
    object. Another is an object that abstracts weights
    $\{W_t^{(i)}\}_{i=1}^N$. By default this is a \code{WeightSet} object. The
    last is a collection of \rng streams, one for each particle. By default
    this is an \code{RngSet} object.
  \item Operations that perform tasks common to all samplers to these
    particles, such as resampling. These are the member functions of
    \code{Particle<T>}.
  \item A sampler that updates the particles (state values and weights) using
    user defined callbacks. This is a \code{Sampler<T>} object.
  \item Monitors that record the importance sampling estimates of certain
    functions defined for the values when the sampler iterates. These are
    \code{Monitor<T>} objects. A monitor for the estimates of $E[h(X_t)]$
    computes $h(X_t^{(i)})$ for each $i = 1,\dots,N$. The function value
    $h(X_t)$ is allowed to be a vector.
\end{itemize}
Note that within the core module, all operations are applied to
\code{Particle<T>} objects, that is $\{W_t^{(i)},X_t^{(i)}\}_{i=1}^N$, instead
of a single particle. Later we will see how to write operations that can be
applied to individual particles and can be parallelized easily.

\subsubsection{Program structures}

A \vsmc program usually consists of the following tasks.
\begin{itemize}
  \item Define a value collection type \code{T}.
  \item Constructing a \code{Sampler<T>} object.
  \item Configure the behavior of initialization and updating by adding
    callable objects to the sampler object.
  \item Optionally add monitors.
  \item Initialize and iterate the sampler.
  \item Retrieve the outputs, estimates and other informations.
\end{itemize}
In this section we document how to implement each of these tasks. Within the
\vsmc library, all public classes, namespaces and free functions, are declared
in the namespace \code{vsmc}.

\subsubsection{The value collection}

The template parameter \code{T} is a user defined type that abstracts the
value collection. \vsmc does not restrict how the values shall be actually
stored. They can be stored in memory, spread among nodes of a cluster, in \gpu
memory or even in a database. However this kind of flexibility comes with a
small price. The value collection does need to fulfill two requirements. We
will see later that for most common usage, \vsmc provides readily usable
implementations, on top of which users can create problem specific classes.

First, the value collection class \code{T} has to provide a constructor of the
form
\begin{Code}
T (SizeType N)
\end{Code}
where \code{SizeType} is some integer type. Since \vsmc allows one to allocate
the states in any way suitable, one needs to provide this constructor which
\code{Particle<T>} can use to allocate them.

Second, the class has to provide a member function named \code{copy} that
copies each particle according to replication numbers given by a resampling
algorithm. For the same reason as above \vsmc has no way to know how it can
extract and copy a single particle when it is doing the resampling. The
signature of this member function may look like
\begin{Code}
template <typename SizeType>
void copy (std::size_t N, const SizeType *copy_from);
\end{Code}
The pointer \code{copy_from} points to an array that has $N$ elements,
where $N$ is the number of particles. After calling this member function, the
value of particle \code{i} shall be copied from the particle
\code{j = copy_from[i]}. In other words, particle \code{i} is a child of
particle \code{copy_from[i]}, or \code{copy_from[i]} is the parent of particle
\code{i}. If a particle \code{j} shall remain itself, then
\code{copy_from[j] == j}. How the values are actually copied is user defined.
Note that, the member function can take other forms, as usual when using \cpp
template generic programming.
The actual type of the pointer \code{copy_from}, \code{SizeType}, is
\code{Particle<T>::size_type}, which depends on the type \code{T}. For
example, define the member function as the following is also allowed,
\begin{Code}
void copy (int N, const std::size_t *copy_from);
\end{Code}
provided that \code{Particle<T>::size_type} is indeed \code{std::size_t}.
However, writing it as a member function template releases the users from
finding the actual type of pointer \code{copy_from} and the sometimes
troubling forward declaration issues. Will not elaborate such more technical
issues further.

\subsubsection{Constructing a sampler}

Once the value collection class \code{T} is defined. One can start
constructing \smc samplers. For example, the following line creates a sampler
with $N$ particles
\begin{Code}
Sampler<T> sampler(N);
\end{Code}
The number of particles is the only mandatory argument of the constructor.
There are two optional parameters. The complete signature of the constructor
is,
\begin{Code}
explicit Sampler<T>::Sampler (Sampler<T>::size_type N,
        ResampleScheme scheme = Stratified, double threshold = 0.5);
\end{Code}
The \code{scheme} parameter is self-explanatory. \vsmc provides a few built-in
resampling schemes; see the reference manual for a list of them. User defined
resampling algorithms can also be used. See the reference manual for details.
The \code{threshold} is the threshold of $\ess/N$ below which a resampling
will be performed. It is obvious that if $\text{\code{threshold}}\ge1$ then
resampling will be performed at each iteration. If
$\text{\code{threshold}}\le0$ then resampling will never be performed. Both
parameters can be changed later. However the size of the sampler can never be
changed after the construction.

\subsubsection{Initialization and updating}

All the callable objects that initialize, move and weight the particle
collection can be added to a sampler through a set of member functions. All
these objects operate on the \code{Particle<T>} object. Because \vsmc
allows one to manipulate the particle collection as a whole, in principle many
kinds of parallelization are possible.

To set an initialization method, one need to implement a function with the
following signature,
\begin{Code}
std::size_t init_func (Particle<T> &particle, void *param);
\end{Code}
or a class with \code{operator()} properly defined, such as
\begin{Code}
struct init_class
{ std::size_t operator() (Particle<T> &particle, void *param); };
\end{Code}
They can be added to the sampler through
\begin{Code}
sampler.init(init_func);
\end{Code}
or
\begin{Code}
sampler.init(init_class());
\end{Code}
respectively. \cppoo{} \code{std::function} or its \boost equivalent
\code{boost::function} can also be used. For example,
\begin{Code}
std::function<std::size_t (Particle<T> &, void *)> init_obj(init_func);
sampler.init(init_obj);
\end{Code}

The addition of updating methods is more flexible. There are two kinds of
updating methods. One is simply called \code{move} in \vsmc, and is performed
before the possible resampling at each iteration. These moves usually perform
the updating of the weights among other tasks. The other is called
\code{mcmc}, and is performed after the possible resampling. They are often
\mcmc type moves. Multiple \code{move}'s or \code{mcmc}'s are also allowed. In
fact a \vsmc sampler consists of a queue of \code{move}'s and a queue of
\code{mcmc}'s. The \code{move}'s in the queue can be changed through
\code{Sampler<T>::move},
\begin{Code}
Sampler<T> &move (const Sampler<T>::move_type &new_move, bool append);
\end{Code}
If \code{append == true} then \code{new_move} is appended to the existing
(possibly empty) queue. Otherwise, the existing queue is cleared and
\code{new_move} is added. The member function returns a reference to the
updated sampler. For example, the following move,
\begin{Code}
std::size_t move_func (std::size_t iter, Particle<T> &particle);
\end{Code}
can be added to a sampler by
\begin{Code}
sampler.move(move_func, false);
\end{Code}
This will clear the (possibly empty) existing queue of \code{move}'s and set a
new one. To add multiple moves into the queue,
\begin{Code}
sampler.move(move1, true).move(move2, true).move(move3, true);
\end{Code}
Objects of class type with \code{operator()} properly defined can also be
used, similarly to the initialization method. The queue of \code{mcmc}'s can
be used similarly. See the reference manual for other methods that can be used
to manipulate these two queues.

In principle, one can combine all moves into a single move. However, sometimes
it is more natural to think of a queue of moves. For instance, if a
multi-block Metropolis random walk consists of kernels $K_1$ and $K_2$, then
one can implement each of them as functions, say \code{mcmc_k1} and
\code{mcmc_k2}, and add them to the sampler sequentially,
\begin{Code}
sampler.mcmc(mcmc_k1, true).mcmc(mcmc_k2, true);
\end{Code}
Then at each iteration, they will be applied to the particle collection
sequentially in the order in which they are added.

\subsubsection{Running the algorithm, monitoring and outputs}

Having set all the operations, one can initialize and iterate the sampler by
calling
\begin{Code}
sampler.initialize((void *)param);
sampler.iterate(iter_num);
\end{Code}
The \code{param} argument to \code{initialize} is optional, with \code{NULL}
as its default. This parameter is passed to the user defined \code{init_func}.
The \code{iter_num} argument to \code{iterate} is also optional; the default
is $1$.

Before initializing the sampler or after a certain time point, one can add
monitors to the sampler. The concept is similar to \cbugs's \code{monitor}
statement, except it does not monitor the individual values but rather the
importance sampling estimates. Consider approximating $\Exp[h(X_t)]$, where
$h(X_t) = (h_1(X_t),\dots,h_m(X_t))$ is an $m$-vector function. The importance
sampling estimate can be obtained by $AW$ where $A$ is an $N$ by $m$ matrix
where $A(i,j) = h_j(X_t^{(i)})$ and $W = (W_t^{(i)},\dots,W_t^{(N)})^T$ is the
$N$-vector of normalized weights. To compute this importance sampling
estimate, one need to define the following evaluation function (or a class
with \code{operator()} properly defined),
\begin{Code}
void monitor_eval (std::size_t iter, std::size_t m,
        const Particle<T> &particle, double *res)
\end{Code}
and add it to the sampler by calling,
\begin{Code}
sampler.monitor("some.user.chosen.variable.name", m, monitor_eval);
\end{Code}
When the function \code{monitor_eval} is called, \code{iter} is the iteration
number of the sampler, \code{m} is the same value as the one the user passed
to \code{Sampler<T>::monitor}; and thus one does not need global variable or
other similar techniques to access this value. The output pointer \code{res}
points to an $N \times m$ output array of row major order. That is, after the
calling of the function, \code{res[i * dim + j]} shall be $h_j(X_t^{(i)})$.

After each iteration of the sampler, the importance sampling estimate will be
computed automatically. See the reference manual for various ways to retrieve
the results. Usually it is sufficient to output the sampler by
\begin{Code}
std::ofstream output("file.name");
output << sampler << std::endl;
\end{Code}
where the output file will contain the importance sampling estimates among
other things. Alternatively, one can use the \code{Monitor<T>::record} member
function to access specific historical results. See the reference manual for
details of various overloaded version of this member function.

A reference to the value collection \code{T} object can be retrieved through
the \code{Particle<T>::value} member function. \code{Particle<T>} objects
manage the weights through a weight set object, which by default is of type
\code{WeightSet}. The \code{Particle<T>::weight_set} member function returns a
reference to this weigh set object. A user defined weight set class that
abstracts $\{W_t^{(i)}\}_{i=1}^N$ can also be used. The details involve some
more advanced \cpp template techniques and are documented in the reference
manual. One possible reason for replacing the default is to provide special
memory management of the weight set. For example, the \mpi module provides a
special weight set class that manages weights across multiple nodes and
perform proper normalization, computation of \ess, and other tasks.

The default \code{WeightSet} object provides some ways to retrieve weights.
Here we document some of the most commonly used. See the reference manual for
details of others. The weights can be accessed one by one, for example,
\begin{Code}
Particle<T> &particle = sampler.particle();
double w_i     = particle.weight_set().weight(i);
double log_w_i = particle.weight_set().log_weight(i);
\end{Code}
One can also read all weights into a container, for example,
\begin{Code}
std::vector<double> w(particle.size());
particle.weight_set().read_weight(w.begin());
double *lw = new double[particle.size()];
particle.weight_set().read_log_weight(lw);
\end{Code}
Note that these member functions accept general output iterators.

\subsubsection{Implementing initialization and updating}

So far we have only discussed how to add initialization and updating objects
to a sampler. To actually implement them, one writes callable objects that
operate on the \code{Particle<T>} object. For example, a move can be
implemented through the following function as mentioned before,
\begin{Code}
std::size_t move_func (std::size_t iter, Particle<T> &particle);
\end{Code}
Inside the body of this function, one can change the value by manipulating the
object through the reference returned by \code{particle.value()}.

When using the default weight set class, the weights can be updated through a
set of member functions of \code{WeightSet}. For example,
\begin{Code}
std::vector<double> weight(particle.size());
particle.weight_set().set_equal_weight();
particle.weight_set().set_weight(weight.begin());
particle.weight_set().mul_weight(weight.begin());
particle.weight_set().set_log_weight(weight.begin());
particle.weight_set().add_log_weight(weight.begin());
\end{Code}
The \code{set_equal_weight} member function sets all weights to be equal. The
\code{set_weight} and \code{set_log_weight} member functions set the values of
weights and logarithm weights, respectively. The \code{mul_weight} and
\code{add_log_weight} member functions multiply the weights or add to the
logarithm weights by the given values, respectively. All these member functions
accept general input iterators as their arguments.

One important thing to note is that, whenever one of these member functions is
called, both the weights and logarithm weights will be re-calculated,
normalized, and the \ess will be updated. The reason for not allowing changing
a single particle's weight is that, in a multi-threading environment, it is
possible for one to change one weight in one thread, and obtain another in
another thread without proper normalizing. Conceptually, changing one weight
actually changes all weights.

\subsubsection{Generating random numbers}

The \code{Particle<T>} object has a sub-object, a collection of \rng engines
that can be used with \cppoo{} \code{<random>} or \boost distributions. For
each particle \code{i}, one can obtain an engine that produces an \rng stream
independent of others by
\begin{Code}
particle.rng(i);
\end{Code}
To generate distribution random variates, one can use the
\cppoo{} \code{<random>} library. For example,
\begin{Code}
std::normal_distribution<double> rnorm(mean, sd);
double r = rnorm(particle.rng(i));
\end{Code}
or use the \boost library,
\begin{Code}
boost::random::normal_distribution<double> rnorm(mean, sd);
double r = rnorm(particle.rng(i));
\end{Code}
\vsmc itself also makes use of \cppoo{} \code{<random>} or \boost depending
on the value of the configuration macro \code{VSMC_HAS_CXX11LIB_RANDOM}.
Though the user is free to choose which one to use in their own code, there is
a convenient alternative. For each class defined in \cppoo{} \code{<random>},
it is imported to the \code{vsmc::cxx11} namespace. Therefore one can use
\begin{Code}
cxx11::normal_distribution<double> rnorm(mean, sd);
\end{Code}
while the underlying implementation of \code{normal_distribution} can be
either \cppoo standard library or \boost. The benefit is that if one needs to
develop on multiple platforms, and only some of them support \cppoo and some
of them have the \boost library, then only the configure macro
\code{VSMC_HAS_CXX11LIB_RANDOM} needs to be changed. This can be configured
through \cmake and other build systems. Of course, one can also use an
entirely different \rng system than those provided by \vsmc.

\subsection{SMP module}
\label{sub:SMP module}

\subsubsection{The value collection}

Many typical problems' value collections can be viewed as a matrix of certain
type. For example, a simple particle filter whose state is a vector of length
\code{Dim} and type \code{double} can be viewed as an $N$ by \code{Dim} matrix
where $N$ is the number of particles. A trans-dimensional problem can use an
$N$ by $1$ matrix whose type is a user defined class, say \code{StateType}.
For this kind of problems, \vsmc provide a class template
\begin{Code}
template <MatrixOrder Order, std::size_t Dim, typename StateType>
class StateMatrix;
\end{Code}
which provides the constructor and the \code{copy} member function required by
the core module interface, as well as methods for accessing individual values.
The first template parameter (possible value \code{RowMajor} or
\code{ColMajor}) specifies how the values are ordered in memory. Usually one
shall choose \code{RowMajor} to optimize data access. The second template
parameter is the number of variables, an integer value no less than $1$ or the
special value \code{Dynamic}, in which case \code{StateMatrix} provides a
member function \code{resize_dim} such that the number of variables can be
changed at runtime. The third template parameter is the type of the state
values.

Each particle's state is thus a vector of length \code{Dim}, indexed from
\code{0} to \code{Dim - 1}. To obtain the value at position \code{pos} of the
vector of particle \code{i}, one can use one of the following member
functions,
\begin{Code}
StateBase<RowMajor, Dim, StateType> value(N);
StateType val = value.state(i, pos);
StateType val = value.state(i, Position<Pos>());
StateType val = value.state<Pos>(i);
\end{Code}
where \code{Pos} is a compile time constant expression whose value is the same
as \code{pos}, assuming the position is known at compile time. One can also
read all values. To read the variable at position \code{pos},
\begin{Code}
std::vector<StateType> vec(value.size());
value.read_state(pos, vec.begin());
\end{Code}
Or one can read all values through an iterator,
\begin{Code}
std::vector<StateType> mat(Dim * value.size());
value.read_state_matrix(ColMajor, mat.first());
\end{Code}
Alternatively, one can also read all values through an iterator which points
to iterators,
\begin{Code}
std::vector<std::vector<StateType> > mat(Dim);
for (std::size_t i = 0; i != Dim; ++i)
    mat[i].resize(value.size());
std::vector<std::vector<StateType>::iterator> iter(Dim);
for (std::size_t i = 0; i != Dim; ++i)
    iter[i] = mat[i].begin();
value.read_state_matrix(iter.first());
\end{Code}

If the compiler support \cppoo{} \code{<tuple>}, vSMC also provides a
\code{StateTuple} class template, which is similar to \code{StateMatrix}
except that the types of values do not have to be the same for each variable.
This is similar to \crlang's \code{data.frame}. For example, suppose each
particle's state is formed by two \code{double}'s, an \code{int} and a user
defined type \code{StateType}, then the following constructs a value
collection using \code{StateTuple},
\begin{Code}
StateTuple<ColMajor, double, double, int, StateType> value(N);
\end{Code}
And there are a few ways to access the state values, similar to
\code{StateMatrix},
\begin{Code}
double x0 = value.state(i, Position<0>());
int x2 = value.state<2>(i);
std::vector<StateType> vec(value.size());
state.read_state(Position<3>(), vec.begin());
\end{Code}
See the reference manual for details.

\subsubsection{A single particle}

For a \code{Particle<T>} object, one can construct a \code{SingleParticle<T>}
object that abstracts one of the particle from the collection. For example,
\begin{Code}
Particle<T> particle(N);
SingleParticle<T> sp(i, &particle);
\end{Code}
create a \code{SingleParticle<T>} object corresponding to the particle at
position \code{i}. There are a few member functions of
\code{SingleParticle<T>} that makes access to individual particles easier than
through the interface of \code{Particle<T>}. Firt \code{sp.id()} returns the
value of the argument \code{i} in the above code that created this
\code{SingleParticle<T>} object. In addition, \code{sp.rng()} is equivalent to
\code{particle.rng(i)}. Also \code{sp.particle()} returns a constant reference
to the \code{Particle<T>} object. And \code{sp.particle_ptr()} returns a
pointer to such a constant \code{Particle<T>} object. Note that, one cannot
get write access to a \code{Particle<T>} object through interface of
\code{SingleParticle<T>}. Instead, one can only get write access to a single
particle. For example, If \code{T} is a \code{StateMatrix} instantiation or
its derived class, then \code{sp.state(pos)} is equivalent to
\code{particle.value().state(i, pos)} and the reference it returns is mutable.
See the reference manual for more informations on the interface of the
\code{SingleParticle} class template. \code{SingleParticle<T>} objects are
usually not constructed by users, but rather by the libraries' other classes
in the \smp module, and passed to user defined functions, as we will see very
soon.

\subsubsection{Sequential and parallel implementations}

Once we have the \code{SingleParticle<T>} concept, we are ready to introduce
how to write implementations of \smc algorithms that manipulate a single
particle and can be applied to all particles in parallel or sequentially. For
sequential implementations, this can be done through five base classes,
\begin{Code}
template <typename BaseState> class StateSEQ;
template <typename T, typename D = Virtual> class InitializeSEQ;
template <typename T, typename D = Virtual> class MoveSEQ;
template <typename T, typename D = Virtual> class MonitorEvalSEQ;
template <typename T, typename D = Virtual> class PathEvalSEQ;
\end{Code}
The template parameter \code{BaseState} needs to satisfy the general value
collection requirements in addition to a \code{copy_particle} member function,
for example, \code{StateMatrix}. Other base classes expect \code{T} to satisfy
general value collection requirements. The details of all these class
templates can be found in the reference manual. Here we use the
\code{MoveSEQ<T>} class as an example to illustrate their usage. Recall that
\code{Sampler<T>} expect a callable object which has the following signature
as a move,
\begin{Code}
std::size_t move_func (std::size_t iter, Particle<T> &particle);
\end{Code}
For the purpose of illustration, the type \code{T} is defined as,
\begin{Code}
typedef StateMatrix<RowMajor, 1, double> T;
\end{Code}
Here is a typical example of implementation of such a function,
\begin{Code}
std::size_t move_func (std::size_t iter, Particle<T> &particle)
{
    std::vector<double> inc_weight(particle.size());
    for (Particle<T>::size_type i = 0; i != particle.size(); ++i) {
        cxx11::normal_distribution<double> rnorm(0, 1);
        particle.value().state(i, 0) = rnorm(particle.rng(i));
        inc_weight[i] = cal_inc_weight(particle.value().state(i, 0);
    }
    particle.weight_set().add_log_weight(inc_weight.begin());
}
\end{Code}
where \code{cal_inc_weight} is some function that calculates the logarithm
incremental weights. As we see, there are three main parts of a typical move.
First, we allocate a vector \code{inc_weight}. Second, we generate normal
random variates for each particle and calculate the incremental weights. This
is done through a \code{for} loop. Third, we add the logarithm incremental
weights. The first and the third are \emph{global} operations while the second
is \emph{local}. The first and the third are often optional and absent. The
local operation is also usually the most computational intensive part of \smc
algorithms and can benefit the most from parallelizations.

With \code{MoveSEQ<T>}, one can derive from this class, which defines
the \code{operator()} required by the core module interface,
\begin{Code}
std::size_t operator() (std::size_t iter, Particle<T> &particle);
\end{Code}
and customize what this operator does by defining one or more of the following
three member functions, corresponding to the three parts, respectively,
\begin{Code}
void pre_processor (std::size_t iter, Particle<T> &particle);
std::size_t move_state (std::size_t iter, SingleParticle<T> sp);
void post_processor (std::size_t iter, Particle<T> &particle);
\end{Code}
For example,
\begin{Code}
#include <vsmc/smp/backend_seq.hpp>

class move : public MoveSEQ<T>
{
    public :

    void pre_processor (std::size_t iter, Particle<T> &particle)
    {inc_weight_.resize(particle.size());}

    std::size_t move_state (std::size_t iter, SingleParticle<T> sp)
    {
        cxx11::normal_distribution<double> rnorm(0, 1);
        sp.state(0) = rnorm(sp.rng());
        inc_weight_[sp.id()] = cal_inc_weight(sp.state(0));
    }

    void post_processor (std::size_t iter, Particle<T> &particle)
    {particle.weight_set().add_log_weight(inc_weight_.begin());}

    private :

    std::vector<double> inc_weight_;
};
\end{Code}
The \code{operator()} of \code{MoveSEQ<T>} is equivalent to the single
function implementation as shown before.

In the simplest case, \code{MoveSEQ} only takes away the loop around Part 2.
However if one implement the move in such a way, and then replace
\code{MoveSEQ} with \code{MoveOMP}, the changing of the base class name causes
\vsmc to use \openmp to parallelize the loop. For example, one can declare the
class as
\begin{Code}
#include <vsmc/smp/backend_omp.hpp>
class move : public MoveOMP<T>;
\end{Code}
and use exactly the same implementation as before. Now when
\code{move::operator()} is called, it will be parallelized by \openmp. Other
backends are available in case \openmp is not available. Among them there are
\ccilk and \tbb. In addition to these well known parallelization programming
models, \vsmc also has its own implementation using \cppoo{} \code{<thread>}.
There are other backends documented in the reference manual. To use any of
these parallelization, all one need to do is to change a few base class names.
In practice, one can use conditional compilation, for example, to use a
sequential implementation or a \openmp parallelized one, we can write,
\begin{Code}
#ifdef USE_SEQ
#include <vsmc/smp/backend_seq.hpp>
#define BASE_MOVE MoveSEQ
#endif

#ifdef USE_OMP
#include <vsmc/smp/backend_omp.hpp>
#define BASE_MOVE MoveOMP
#endif

class move : public BASE_MOVE<T>;
\end{Code}
And we can compile the same source into different samplers with
\code{Makefile} rules such as the following,
\begin{Code}
prog-seq : prog.cpp
        $(CXX) $(CXXFLAGS) -DUSE_SEQ -o prog-seq prog.cpp
prog-omp : prog.cpp
        $(CXX) $(CXXFLAGS) -DUSE_omp -o prog-omp prog.cpp
\end{Code}
Or one can configure the source file through a build system such as \cmake,
which can also determine which programming model is available on the system.

\subsubsection{Adapter}

Sometimes, the cumbersome task of writing a class to implement a move and
other operations can out weight the power we gain through object oriented
programming. For example, a simple \code{move_state}-like function is all that
we need to get the work done. In this case, one can create a move through the
\code{MoveAdapter}. For example, say we have implemented
\begin{Code}
std::size_t move_state (std::size_t iter, SingleParticle<T> sp);
\end{Code}
as a function. Then one can create a callable object acceptable to
\code{Sampler<T>::move} through
\begin{Code}
MoveAdapter<T, MoveSEQ>  move_obj(move_state);
MoveAdapter<T, MoveSTD>  move_obj(move_state);
MoveAdapter<T, MoveTBB>  move_obj(move_state);
MoveAdapter<T, MoveCILK> move_obj(move_state);
MoveAdapter<T, MoveOMP>  move_obj(move_state);
sampler.move(move_obj, false);
\end{Code}
These are respectively, sequential, \cppoo{} \code{<thread>}, \tbb, \cilk,
and \openmp implementations. The first template parameter is the type of value
collection and the second is the name of the base class template. Actually,
the \code{MoveAdapter}'s constructor accepts two more optional arguments, the
\code{pre_processor} and the \code{post_processor}, corresponding to the other
two aforementioned member functions. Similar adapters for the other three base
classes also exist.

Another scenario where an adapter is desired is that which backend to use
needs be decided at runtime. The above simple adapters can already be used for
this purpose. In addition, another form of the adapter is as the following,
\begin{Code}
class move;
MoveAdapter<T, MoveTBB, move> move_obj;
sampler.move(move_obj, false);
\end{Code}
where the class \code{move} has the same definition as before but it no longer
derives from any base class. The class \code{move} is required to have a
default constructor, a copy constructor and an assignment operator.

\subsection{Thread-safety and scalability considerations}
\label{sub:Thread-safety and scalability considerations}

When implementing parallelized \smc algorithms, issues such as thread-safety
cannot be avoided even though the \vsmc library hides most parallel constructs
from the user.

Classes in the \vsmc library usually guarantee that their member functions are
thread-safe in the sense that calling the same member function on different
objects at the same time from different threads is safe. However, calling
mutable member functions on the same object from different threads is usually
not safe. Calling immutable member functions is generally safe. There are a
few exceptions,
\begin{itemize}
  \item The constructors of \code{Particle} and \code{Sampler} are not
    thread-safe. Therefore if one need to construct multiple \code{Sampler}
    from different threads, a mutex protection is needed. However, subsequent
    member function calls on the constructed objects are thread-safe according
    to the above rules.
  \item Member functions that concern a single particle are generally
    thread-safe in the sense that one can call them on the same object from
    different threads as long as they are called for different particles. For
    example \code{Particle::rng} and \code{StateMatrix::state} are
    thread-safe.
\end{itemize}
In general, one can safely manipulate different individual particles from
different threads, which is the minimal requirement for scalable
parallelization. But one cannot manipulate the whole particle collection from
different threads, for example \code{WeightSet::set_log_weight}.

User defined callbacks shall generally follow the same rules. For example, for
a \code{MoveOMP} subclass, \code{pre_processor} and \code{post_processor} does
not have be thread-safe, but \code{move_state} needs to be. In general, avoid
write access to memory locations shared by all particles from
\code{move_state} and other similar member functions. One needs to take some
extra care when using third-party libraries. For example, in our example
implementation of the \code{move} class, the \code{rnorm} object, which is
used to generate Normal random variates, is defined within \code{move_state}
instead of being a class member data even though it is created with the same
parameters for each particle. This is because the call \code{rnorm(sp.rng())}
is not thread-safe in some implementations, for example, when using the \boost
library.

For scalable performance, certain practices should be avoided when
implementing member functions such as \code{move_state}. For example, dynamic
memory allocation is usually lock-based and thus should be avoided.
Alternatively one can use a scalable memory allocator such as the one provided
by \tbb. In general, in functions such as \code{move_state}, one should avoid
using locks to guarantee thread-safety, which can be a bottleneck to parallel
performance.

\section{Example}
\label{sec:Example}

\subsection{A simple particle filter}
\label{sub:A simple particle filter}

\subsubsection{The model and algorithm}

This is an example used in \csmctc. Through this example, we will show how to
re-implement a simple particle filter in \vsmc. It shall walk one through the
basic features of the library. \smctc is the first general framework for
implementing \smc algorithms in \cpp and serves as one of the most important
inspirations for the \vsmc library. It is widely used by many researchers. One
of the goals of the current work is that users familiar with \smctc shall find
little difficulty in using the new library.

The state space model, known as the almost constant velocity model in the
tracking literature, provides a simple scenario. The state vector $X_t$
contains the position and velocity of an object moving in a plane. That is,
$X_t = (\xpos^t, \ypos^t, \xvel^t, \yvel^t)^T$. Imperfect observations $Y_t =
(\xobs^t, \yobs^t)^T$ of the positions are possible at each time instance. The
state and observation equations are linear with additive noises,
\begin{align*}
  X_t &= AX_{t-1} + V_t \\
  Y_t &= BX_t + \alpha W_t
\end{align*}
where
\begin{equation*}
  A = \begin{pmatrix}
    1 & \Delta & 0 & 0 \\
    0 & 1 & 0 & 0 \\
    0 & 0 & 1 & 0 \\
    0 & 0 & 0 & 1
  \end{pmatrix} \qquad
  B = \begin{pmatrix}
    1 & 0 & 0 & 0 \\
    0 & 1 & 0 & 0 \\
  \end{pmatrix} \qquad
  \alpha = 0.1
\end{equation*}
and we assume that the elements of the noise vector $V_t$ are independent
Normal with variance $0.02$ and $0.001$ for position and velocity,
respectively. The observation noise, $W_t$ comprises independent, identically
distributed $t$-distributed random variables with degree of freedom $\nu =
10$. The prior at time $0$ corresponds to an axis-aligned Gaussian with
variance $4$ for the position coordinates and $1$ for the velocity
coordinates. The particle filter algorithm is shown in Algorithm~\ref{alg:pf}.

\begin{algorithm}[t]
\begin{algorithmic}
  \hrule\vskip1ex
  \STATE \emph{Initialization}
  \STATE\STATESKIP Set $t\leftarrow0$.
  \STATE\STATESKIP Sample
  $\xpos^{(0,i)},\ypos^{(0,i)}\sim\rnorm(0,4)$ and
  $\xvel^{(0,i)},\yvel^{(0,i)}\sim\rnorm(0,1)$.
  \STATE\STATESKIP Weight $W_0^{(i)} \propto L(X_0^{(i)}|Y_0)$ where $L$ is
  the likelihood function.

  \STATE \emph{Iteration}
  \STATE\STATESKIP Set $t\leftarrow t + 1$.
  \STATE\STATESKIP Sample
  \begin{align*}
    \xpos^{(t,i)}&\sim\rnorm(\xpos^{(t-1,i)} + \Delta\xvel^{(t-1,i)}, 0.02) &
    \xvel^{(t,i)}&\sim\rnorm(\xvel^{(t-1,i)}, 0.001) \\
    \ypos^{(t,i)}&\sim\rnorm(\ypos^{(t-1,i)} + \Delta\yvel^{(t-1,i)}, 0.02) &
    \yvel^{(t,i)}&\sim\rnorm(\yvel^{(t-1,i)}, 0.001)
  \end{align*}
  \STATE\STATESKIP Weight $W_t^{(i)} \propto W_{t-1}^{(i)}L(X_t^{(i)}|Y_t)$.

  \STATE \emph{Repeat the \emph{Iteration} step until all data are processed}.
  \vskip1ex\hrule
\end{algorithmic}
\caption{Particle filter algorithm for the almost constant velocity model.}
\label{alg:pf}
\end{algorithm}

\subsubsection{Implementations}

We first introduce the body of the \code{main} function, showing the typical
work flow of a \vsmc program.
\begin{Code}
#include <vsmc/core/sampler.hpp>
#include <vsmc/smp/adapter.hpp>
#include <vsmc/smp/state_matrix.hpp>
#include <vsmc/smp/backend_seq.hpp>

const std::size_t DataNum = 100;
const std::size_t ParticleNum = 1000;
const std::size_t Dim = 4;

int main ()
{
    Sampler<cv> sampler(ParticleNum);
    sampler.init(cv_init);
    sampler.move(cv_move(), false);
    MonitorEvalAdapter<cv, MonitorEvalSEQ> cv_est(cv_monitor_state);
    sampler.monitor("pos", 2, cv_est);

    sampler.initialize((void *)"pf.data");
    sampler.iterate(DataNum - 1);

    std::ofstream est("pf.est");
    est << sampler << std::endl;
    est.close();
    est.clear();

    return 0;
}
\end{Code}
In the \code{main} function, we constructed a sampler with \code{ParticleNum}
particles. And then we added the initialization function \code{cv_init} and
the move object of type \code{cv_move}. And then we added a monitor that will
record the importance sampling estimates of the two position parameters. Next,
we initialized the sampler with data file \code{pf.data} and iterate the
sampler until all data are processed. The last step is that we output the
results into a file called \code{pf.est}.

The class \code{cv} will be our value collection which is a subclass of
\code{StateMatrix<RowMajor, Dim, double>}. To illustrate both the core module
and the \smp module, the initialization \code{cv_init} will be implemented
as a standalone function. The move \code{cv_move} will be implemented as a
derived class of \code{MoveSEQ<cv>}. To monitor the importance sampling
estimates of the two position parameters, we will implement a simple function
\code{cv_monitor_state} and use the adapter \code{MonitorEvalAdapter<cv,
  MonitorEvalSEQ>}.

The value collection is an $N$ by \code{Dim} (in this case $\text{\code{Dim}}
= 4$) matrix of type \code{double}. We can simply use
\code{StateMatrix<RowMajor, Dim, double>} as our value collection. However, we
would like to enhance its functionality through inheritance. First, since the
data is shared by all particles, it is natural to bind it with the value
collection. Second, both the initialization and move will need to calculate
the log-likelihood. We can implement it as a standalone function, but since
the log-likelihood function will need to access the data, it is convenient to
implement it as a member function of the value collection \code{cv}. Here is
the full implementation of this simple value collection class
\begin{Code}
class cv : public StateMatrix<RowMajor, Dim, double>
{
    public :

    cv (size_type N) :
        StateMatrix<RowMajor, Dim, double>(N),
        x_obs_(DataNum), y_obs_(DataNum) {}

    double log_likelihood (std::size_t iter, size_type id) const
    {
        const double scale = 10;
        const double nu = 10;
        double llh_x = scale * (state(id, 0) - x_obs_[iter]);
        double llh_y = scale * (state(id, 1) - y_obs_[iter]);
        llh_x = std::log(1 + llh_x * llh_x / nu);
        llh_y = std::log(1 + llh_y * llh_y / nu);

        return -0.5 * (nu + 1) * (llh_x + llh_y);
    }

    void read_data (const char *filename)
    {
        std::ifstream data(filename);
        for (std::size_t i = 0; i != DataNum; ++i)
            data >> x_obs_[i] >> y_obs_[i];
        data.close();
        data.clear();
    }

    private :

    std::vector<double> x_obs_;
    std::vector<double> y_obs_;
};
\end{Code}
The \code{log_likelihood} member function accepts the iteration number and the
particle's \code{id} as arguments. It returns the log-likelihood of the
\code{id}'th particle at iteration \code{iter}. The \code{read_data} member
function simply read the data from a file.

The initialization is implemented through the \code{cv_init} function,
\begin{Code}
std::size_t cv_init (Particle<cv> &particle, void *filename);
\end{Code}
such that, it first checks if \code{filename} is \code{NULL}. If it is not,
then we use it to read the data. So the first initialization may look like
\begin{Code}
sampler.initialize((void *)filename);
\end{Code}
And after that, if we want to re-initialize the sampler, we can simply call,
\begin{Code}
sampler.initialize();
\end{Code}
This will reset the sampler and initialize it again but without reading the
data. If the data set is large, repeated \io can be very expensive. After
reading the data, we will initialize each particle's value by Normal random
variates, and calculate its log-likelihood. The last step is to set the
logarithm weights of the particle collection. Since this is not a
accept-reject type algorithm, the returned acceptance count bares no meaning.
Here is the complete implementation,
\begin{Code}
std::size_t cv_init (Particle<cv> &particle, void *filename)
{
    if (filename)
        particle.value().read_data(static_cast<const char *>(filename));

    const double sd_pos0 = 2;
    const double sd_vel0 = 1;
    cxx11::normal_distribution<double> norm_pos(0, sd_pos0);
    cxx11::normal_distribution<double> norm_vel(0, sd_vel0);
    std::vector<double> log_weight(particle.size());

    for (Particle<cv>::size_type i = 0; i != particle.size(); ++i) {
        particle.value().state(i, 0) = norm_pos(particle.rng(i));
        particle.value().state(i, 1) = norm_pos(particle.rng(i));
        particle.value().state(i, 2) = norm_vel(particle.rng(i));
        particle.value().state(i, 3) = norm_vel(particle.rng(i));
        log_weight[i] = particle.value().log_likelihood(0,i);
    }
    particle.weight_set().set_log_weight(log_weight.begin());

    return 0;
}
\end{Code}
In this example, we read all data from a single file for simplicity. In a
realistic application, the data is often processed online -- the filter is
applied when new data becomes available. In this case, the user can use the
optional argument of \code{Sampler<T>::initialize} to pass necessary
information to open a data connection instead of a file name.

In the above implementation we iterated over all particles. There are other
ways to iterate over a particle collection. First we can use
\code{SingleParticle<cv>} objects,
\begin{Code}
for (Particle<cv>::size_type i = 0; i != particle.size(); ++i) {
    SingleParticle<cv> sp(i, &particle);
    sp.state(0) = norm_pos(sp.rng());
}
\end{Code}
Second, \vsmc provides \stl style iterators, \code{Particle<T>::iterator},
whose \code{value_type} is \code{SingleParticle<T>}. Therefore one can write
the following loop
\begin{Code}
for (Particle<cv>::iterator iter = particle.begin();
     iter != particle.end(); ++iter)
    iter->state(0) = norm_pos(iter->rng());
\end{Code}
Third, if one has a compiler supporting \cppoo{} \code{auto} and range-based
\code{for}, the following is also supported,
\begin{Code}
for (auto &sp : particle) sp.state(0) = norm_pos(sp.rng());
\end{Code}
There are little or no performance difference among all these forms of loops.
However, one can choose an appropriate form to work with the interfaces of
other libraries. For example, the iterator support allows \vsmc to be used
with \stl{} \code{<algorithm>} and other similar libraries.

The updating method \code{cv_move} is similar to \code{cv_init}. It will
update the particle's value by adding Normal random variates. However, as we
see above, each call to \code{cv_init} causes a \code{log_weight} vector being
allocated. Its size does not change between iterations. So it can be viewed as
some resource of \code{cv_move} and it is natural to use a class object to
manage it. Here is the implementation of \code{cv_move},
\begin{Code}
class cv_move : public MoveSEQ<cv>
{
    public :

    void pre_processor (std::size_t iter, Particle<cv> &particle)
    {incw_.resize(particle.size());}

    std::size_t move_state (std::size_t iter, SingleParticle<cv> sp)
    {
        const double sd_pos = std::sqrt(0.02);
        const double sd_vel = std::sqrt(0.001);
        const double delta = 0.1;
        cxx11::normal_distribution<double> norm_pos(0, sd_pos);
        cxx11::normal_distribution<double> norm_vel(0, sd_vel);

        sp.state(0) += norm_pos(sp.rng()) + delta * sp.state(2);
        sp.state(1) += norm_pos(sp.rng()) + delta * sp.state(3);
        sp.state(2) += norm_vel(sp.rng());
        sp.state(3) += norm_vel(sp.rng());

        incw_[sp.id()] = sp.particle().value().log_likelihood(iter, sp.id());

        return 0;
    }

    void post_processor (std::size_t iter, Particle<cv> &particle)
    {particle.weight_set().add_log_weight(&incw_[0]);}

    private :

    std::vector<double> incw_;
};
\end{Code}
First before calling any \code{move_state}, the \code{pre_processor} will be
called, as described in Section~\ref{sub:SMP module}. At this step, we will
resize the vector used for storing incremental weights. After the first
resizing, subsequent calls to \code{resize} will only cause reallocation if
the size changed. In our example, the size of the particle system is fixed, so
we don't need to worry about excessive dynamic memory allocations. The
\code{move_state} member function moves each particle according to our model.
And after \code{move_state} is called for each particle, the
\code{post_processor} will be called and we simply add the logarithm
incremental weights.

For each particle, we want to monitor the $\xpos$ and $\ypos$ parameters and
get the importance sampling estimates. To extract the two values from a
particle, we can implement the following function
\begin{Code}
void cv_monitor_state (std::size_t iter, std::size_t dim,
    ConstSingleParticle<cv> csp, double *res)
{
    assert(dim <= Dim);
    for (std::size_t d = 0; d != dim; ++d)
        res[d] = csp.state(d);
}
\end{Code}
and in the \code{main} function we construct a monitor by
\begin{Code}
MonitorEvalAdapter<cv, MonitorEvalSEQ> cv_est(cv_monitor_state);
\end{Code}
and add it to the sampler through
\begin{Code}
sampler.monitor("pos", 2, cv_est);
\end{Code}
If later we decided to monitor all states, we only need to change the \code{2}
in the above line to \code{Dim}.

\begin{figure}
  \centering
  \includegraphics[width=\linewidth]{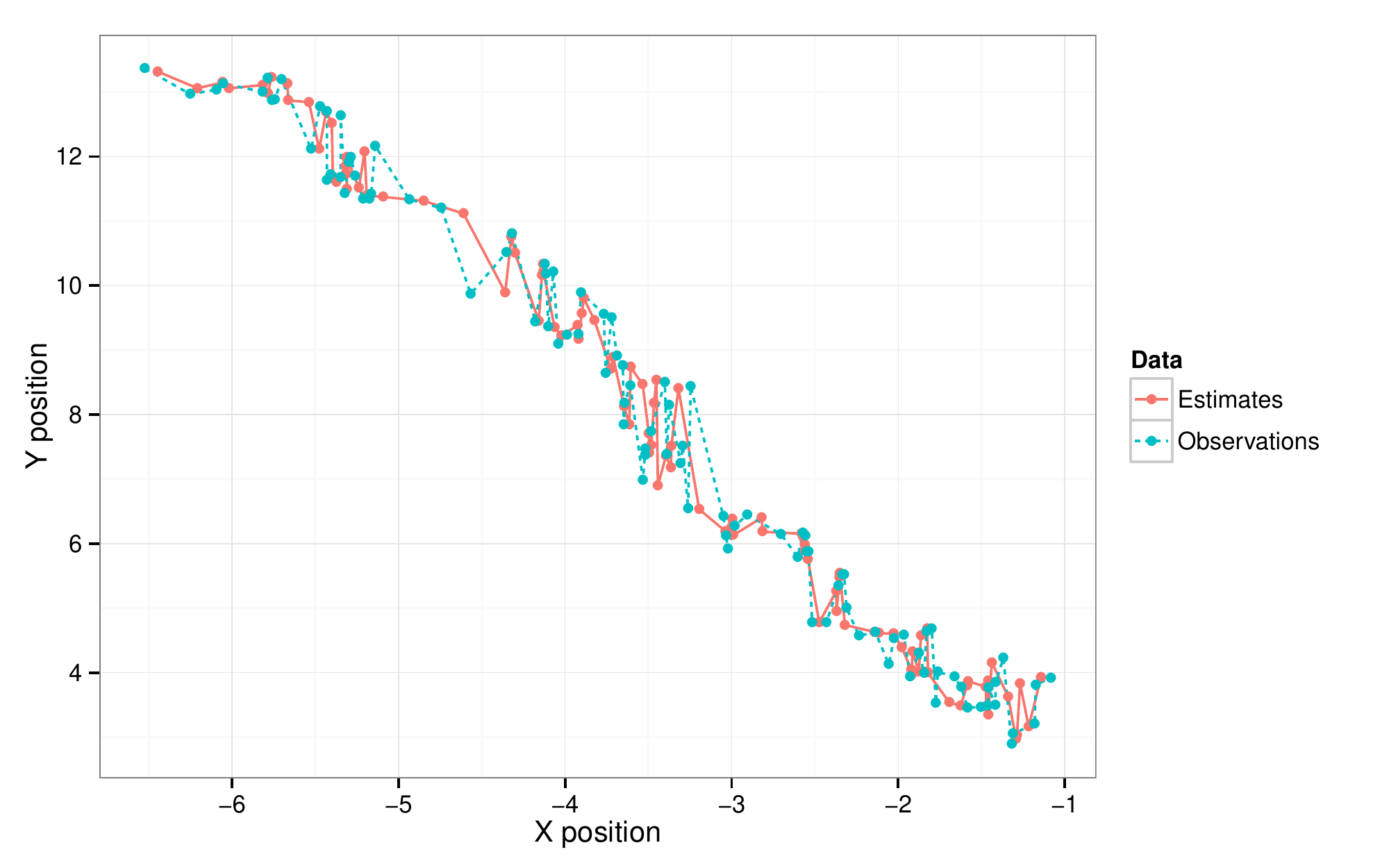}
  \caption{Observations and estimates of a simple particle filter.}
  \label{fig:pf}
\end{figure}

After we implemented all the above, compiled and ran the program, a file
called \code{pf.est} was written by the following statement in the \code{main}
function,
\begin{Code}
est << sampler << std::endl;
\end{Code}
The output file contains the \ess, resampling, importance sampling estimates
and other informations in a table. This file can be read by most statistical
softwares for further analysis. For instance, we can process this file with
\rlang, and get the plot of the estimates of positions against the
observations, as shown in Figure~\ref{fig:pf}.

\subsection{Bayesian Modeling of Gaussian mixture model}
\label{sub:Bayesian Modeling of Gaussian mixture model}

\subsubsection{The model and algorithm}

Since \citet{Richardson:1997ea}, the Gaussian mixture model (\gmm) has
provided a canonical example of a model-order-determination problem. We use
the same model as in \citet{DelMoral:2006hc} to illustrate the implementation
of this classical example in Monte Carlo literature. This model is also used
in \citet{Zhou2013mc} for demonstration of the use of \smc in the context of
Bayesian model comparison, which provides more details of the following
setting. The model is as follows; data $\data = (y_1,\dots,y_n)$ are
independently and identically distributed as
\begin{equation*}
  y_i|\theta_r \sim \sum_{j=1}^r \omega_j\rnorm(\mu_j,\lambda_j^{-1})
\end{equation*}
where $\rnorm(\mu_j,\lambda_j^{-1})$ denotes the Normal distribution with mean
$\mu_j$ and precision $\lambda_j$; $\theta_r =
(\mu_{1:r},\lambda_{1:r},\omega_{1:r})$ and $r$ is the number of components in
each model. The parameter space is thus $\Real^r\times\Real^{+r}\times S_r$
where $S_r = \{\omega_{1:r}:0\le\omega_j\le1; \sum_{j=1}^r\omega_j=1\}$ is the
standard $(r-1)$-simplex. The priors which are the same for each component are
taken to be $\mu_j\sim\rnorm(\xi,\kappa^{-1})$,
$\lambda_j\sim\rgamma(\nu,\chi)$ and $\omega_{1:r}\sim\rdir(\rho)$ where
$\rdir(\rho)$ is the symmetric Dirichlet distribution with parameter $\rho$
and $\rgamma(\nu,\chi)$ is the Gamma distribution with shape $\nu$ and scale
$\chi$. The prior parameters are set in the same manner as in
\citet{Richardson:1997ea}; also see \citet{Zhou2013mc} for details. The data
is simulated from a four components model with $\mu_{1:4} = (-3, 0,3, 6)$, and
$\lambda_j =2$, $\omega_j = 0.25$, $j = 1,\dots,4$. Our interest is to
simulate the posterior distribution of models with $r$ components, denoted by
$M_r$ and obtaining the normalizing constant for the purpose of Bayesian model
comparison \citep[chap.~7]{Robert:2007tc}.

Numerous strategies are possible to construct a sequence of distributions for
the purpose of \smc sampling. One option is to use for each model $M_r$,
$r\in\{1,2,\dots\}$, the sequence $\{\pi_t\}_{t=0}^{T_r}$, defined by
\begin{equation}
  \pi_t(\theta_r^t) \propto
  \pi(\theta_r^t|M_r)p(\data|\theta_r^t,M_r)^{\alpha(t/T_r)}.
  \label{eq:geometry}
\end{equation}
where the number of distribution, $T_r$, and the annealing schedule,
$\alpha:[0,1]\to[0,1]$ may be different for each model. This leads to
Algorithm~\ref{alg:gmm}.

\begin{algorithm}[t]
\begin{algorithmic}
  \hrule\vskip1ex
  \STATE For each model $M_r\in\calM$ perform the following algorithm.

  \STATE \emph{Initialization}
  \STATE\STATESKIP Set $t\leftarrow0$.
  \STATE\STATESKIP Sample $\theta_r^{(i,t)}\sim\pi(\theta_r^{(i,t)}|M_r)$.
  \STATE\STATESKIP Weight $W_0^{(i)} \propto 1$.

  \STATE \emph{Iteration}
  \STATE\STATESKIP Set $t\leftarrow t + 1$.
  \STATE\STATESKIP Weight $W_t^{(i)} \propto W_{t-1}^{(i)}
  p(\data|\theta_r^{(i,t-1)},M_r)^{\alpha(t/T_r) - \alpha([t-1]/T_r)}$.
  \STATE\STATESKIP Apply resampling if necessary.
  \STATE\STATESKIP Sample $\theta_r^{(i,t)} \sim
  K_t(\cdot|\theta_r^{(i,t-1)})$, a $\pi_t$-invariant \mcmc kernel.

  \STATE \emph{Repeat the \emph{Iteration} step up to $t = T_r$}.
  \vskip1ex\hrule
\end{algorithmic}
\caption{\smc algorithm for Bayesian modeling of Gaussian mixture
  model.}
\label{alg:gmm}
\end{algorithm}

The \mcmc kernel $K_t$ in Algorithm~\ref{alg:gmm} is constructed as a
three-blocks Metropolis random walk,
\begin{enumerate}
  \item Update $\mu_{1:r}$ through a Normal random walk.
  \item Update $\lambda_{1:r}$ through a Normal random walk on logarithm
    scale, that is, on $\log\lambda_{j}$, $j = 1, \dots, r$.
  \item Update $\omega_{1:r}$ through a Normal random walk on logit scale,
    that is, on $\omega_{j}/\omega_r$, $j = 1,\dots,r-1$.
\end{enumerate}

The standard direct estimate of the normalizing constants
\citep{DelMoral:2006hc} can be obtained from the output of this \smc
algorithm,
\begin{equation}
  \hat\lambda_{\text{\textsc{ds}}}^{T_r,N} =
  \sum_{i=1}^N \frac{\pi(\theta_r^{(i,0)}|M_r)}{\nu(\theta_0^{(i,0)})} \times
  \prod_{t=1}^{T_r} \sum_{i=1}^N W_{t-1}^{(i)}
  p(\data|\theta_r^{(i,t)}M_r)^{\alpha(t/T_r) - \alpha([t-1]/T_r)}
  \label{eq:smc-ds}
\end{equation}
where $W_{t-1}^{(i)}$ is the importance weight of sample $\theta_{t-1}^{(i)}$.

\subsubsection{Path sampling for estimation of normalizing constants}

As shown in \citet{Zhou2013mc} the estimation of the normalizing constant
associated with our sequence of distributions can also be achieved by a Monte
Carlo approximation to the \emph{path sampling} formulation given by
\citet{Gelman:1998ei}, also known as thermodynamic integration or Ogata's
method. Given a parameter $\alpha$ which defines a family of distributions,
$\{p_{\alpha} = q_{\alpha} / Z_\alpha\}_{\alpha \in [0,1]}$ that move smoothly
from $p_0 = q_0 / Z_0$ to $p_1 = q_1 / Z_1$ as $\alpha$ increases from zero to
one, one can estimate the logarithm of the ratio of their normalizing
constants via a simple integral relationship,
\begin{equation}
  \log\biggl( \frac{Z_1}{Z_0} \biggr) =
  \int_{0}^{1} \Exp_\alpha \biggl[ \rnd{\log q_{\alpha}(\cdot)}{\alpha}
  \biggr] \intd\alpha, \label{eq:path_identity}
\end{equation}
where $\Exp_\alpha$ denotes expectation under $p_\alpha$. The sequence of
distributions in the \smc algorithm for this example can be interpreted as
belonging to such a family of distributions, with $\alpha = \alpha(t/T_r)$.

The \smc sampler provides us with a set of weighted samples obtained from a
sequence of distributions suitable for approximating this integral. At each
time $t$ we can obtain an estimate of the expectation within the integral via
the usual importance sampling estimator, and this integral can then be
approximated via a trapezoidal integration. In summary, the path sampling
estimator of the ratio of normalizing constants $\lambda^{T_r} = \log(Z_1/Z_0)$
can be approximated by
\begin{equation}
  \hat\lambda_{\text{\textsc{ps}}}^{T_r,N} = \sum_{t=1}^{T_r}
  \frac{1}{2}(\alpha_t - \alpha_{t - 1})(U_t^N + U_{t-1}^N)
  \label{eq:path_est}
\end{equation}
where
\begin{equation}
  U_t^N = \sum_{i=1}^N
  W_t^{(i)} \rnd{\log q_{\alpha}(X_t^{(i)})}{\alpha}\Bigm|_{\alpha = \alpha_t}
  \label{eq:path_import}
\end{equation}

\subsubsection{Implementations}

In this example we will implement the following classes.
\begin{itemize}
  \item \code{gmm_param} is a class that abstracts the parameters of the
    model, $\theta_r = (\mu_{1:r},\lambda_{1:r},\omega_{1:r})$.
  \item \code{gmm_state} is the value collection class.
  \item \code{gmm_init} is a class that implements operations used to
    initialize the sampler.
  \item \code{gmm_move_smc} is a class that implements operations used to
    update the weights as well as selecting the random walk proposal scales
    and distribution parameter $\alpha(t/T_r)$.
  \item \code{gmm_move_mu}, \code{gmm_move_lambda} and \code{gmm_move_weight}
    are classes that implement the random walks, each for one of the three
    blocks.
  \item \code{gmm_path} is a class that implements monitors for the path
    sampling estimator. This class is similar to the importance sampling
    monitor introduced before. It is to be used with
    \code{Sampler<gmm_state>::path_sampling}. Its interface requirement will
    be documented later.
  \item \code{gmm_alpha_linear} and \code{gmm_alpha_prior} are classes that
    implement two of the many possible annealing schemes, $\alpha(t/T_r) =
    t/T_r$ (linear) and $\alpha(t/T_r) = (t/T_r)^p$, $p > 1$ (prior).
  \item And last, the \code{main} function, which configure, initialize and
    iterate the samplers.
\end{itemize}

This example is considerably more complicated than the last one. Instead of
documenting all the implementation details, for many classes we will only
show the interfaces. In most cases, the implementations are straightforward as
they are either data member accessors or straight translation of mathematical
formulations. For member functions with more complex structures, detailed
explanation will be given. Interested readers can see the source code for more
details.

Later we will build both sequential and parallelized samplers. A few
configuration macros will be defined at compile time. For example, the
sequential sampler is compiled with the following header and macros,
\begin{Code}
#include <vsmc/smp/backend_seq.hpp>
#define BASE_SATE StateSEQ
#define BASE_INIT InitializeSEQ
#define BASE_MOVE MoveSEQ
#define BASE_PATH PathEvalSEQ
\end{Code}
The definitions of these macros will be changed at compile time to build
parallelized samplers. For example, when using \openmp parallelization, the
header \code{backend_omp.hpp} will be used instead of \code{backend_seq.hpp};
and \code{StateSEQ} will be changed to \code{StateOMP} along with similar
changes to the other macros. In the distributed source code, this is
configured by the \cmake build system.

Again, we first introduce the \code{main} function. The required headers are
the same as the last particle filter example in addition to the \smp backend
headers as described above. The following variables used in the \code{main}
function will be set by the user input.
\begin{Code}
int ParticleNum;
int AnnealingScheme;
int PriorPower;
int CompNum;
std::string DataFile;
\end{Code}
In the \code{main} function, we will create objects that set the distribution
parameter $\alpha(t/T_r)$ at each iteration according to the user input of
\code{AnnealingScheme}. Below is the \code{main} function. Note that some
source code of \io operations which set the parameters above are omitted.
\begin{Code}
int main ()
{
    gmm_move_smc::alpha_setter_type alpha_setter;
    if (AnnealingScheme == 1)
        alpha_setter = gmm_alpha_linear(IterNum);
    if (AnnealingScheme == 2)
        alpha_setter = gmm_alpha_prior(IterNum, PriorPower);

    Sampler<gmm_state> sampler(ParticleNum);
    sampler.particle().value().comp_num(CompNum);
    sampler
        .init(gmm_init());
        .move(gmm_move_smc(alpha_setter), false);
        .mcmc(gmm_move_mu(), false);
        .mcmc(gmm_move_lambda(), true);
        .mcmc(gmm_move_weight(), true);
        .path_sampling(gmm_path());
        .initialize((void *) DataFile.c_str());
        .iterate(IterNum);

    double ds = sampler.particle().value().nc().log_zconst();
    double ps = sampler.path().log_zconst();
    std::cout << "Standard estimate :      " << ds << std::endl;
    std::cout << "Path sampling estimate : " << ps << std::endl;

    return 0;
}
\end{Code}
The sampler first sets the number of components and allocate memory through
member function \code{comp_num} of \code{gmm_state}. Then it sets the
initialization and updating methods. Before possible resampling, a
\code{gmm_move_smc} object is added. After that, three Metropolis random walks
are appended. In addition, we add a \code{gmm_path} object to calculate the
path sampling integration. Then we initialize and iterate the sampler and get
the normalizing constant estimates.

It is obvious that the parameter class \code{gmm_param} need to store the
parameters $(\mu_{1:r},\lambda_{1:r},\omega_{1:r})$. We also associate with
each particle its log-likelihood and log-prior. Here is the definition of the
\code{gmm_param} class. We omitted definitions of some data access member
functions.
\begin{Code}
class gmm_param
{
    public :

    void comp_num (std::size_t num);
    void save_old ();

    double log_prior () const {return log_prior_;}
    double &log_prior () {return log_prior_;}

    double log_likelihood () const {return log_likelihood_;}
    double &log_likelihood () {return log_likelihood_;}

    int mh_reject_mu (double p, double u);
    int mh_reject_lambda (double p, double u);
    int mh_reject_weight (double p, double u);
    int mh_reject_common (double p, double u);

    double log_lambda_diff () const;
    double logit_weight_diff () const;

    void update_log_lambda ();

    private :

    std::size_t comp_num_;
    double log_prior_, log_prior_old_, log_likelihood_, log_likelihood_old_;
    std::vector<double> mu_, mu_old_;
    std::vector<double> lambda_, lambda_old_;
    std::vector<double> weight_, weight_old_;
    std::vector<double> log_lambda_;
};
\end{Code}
The \code{comp_num} member function allocate the memory for a given number of
components. The \code{save_old} member function save the current particle
states. It is used before the states are updated with the random walk
proposals, as we will see later when we implement the \code{gmm_move_mu}. The
\code{mh_reject_mu} member function accept the Metropolis acceptance
probability $p$ and a uniform $(0,1]$ random variate, say $u$; it rejects the
proposed change if $p < u$, and restore the particle state of the parameters
$\mu_{1:r}$ by those values saved by \code{save_old}. The member functions
\code{mh_reject_lambda} and \code{mh_reject_weight} do the same for the other
two set of parameters. All these three also call the \code{mh_reject_common}
which restore the stored log-likelihood and log-prior values. The use of these
member functions will be seen in the implementation of \code{gmm_move_mu}, in
the context of which their own implementation become obvious. Other member
functions provide some useful computations such as the logarithm of the
$\lambda_{1:r}$. They are used when compute the log-likelihood.

The class \code{gmm_state} contains some properties common to all particles,
such as the data and the distribution parameter $\alpha(t/T_r)$. Also, we will
have it to record the logarithm of the ratio of normalizing constants, using
the \code{NormalizingConstant} class. We will see how to update this variable
at each iteration in the implementation of \code{gmm_move_smc}. The prior
parameters are also stored in the value collection. Here is the definition of
this value collection class. Again, we omitted some data access member
functions,
\begin{Code}
class gmm_state : public BASE_STATE<StateMatrix<RowMajor, 1, gmm_param> >
{
    public :

    NormalizingConstant &nc () {return nc_;}
    const NormalizingConstant &nc () const {return nc_;}

    void alpha (double a)
    {
        a = a < 1 ? a : 1;
        a = a > 0 ? a : 0;
        if (a == 0) {
            alpha_inc_ = 0;
            alpha_ = 0;
        } else {
            alpha_inc_ = a - alpha_;
            alpha_ = a;
        }
    }

    void comp_num (std::size_t num)
    {
        comp_num_ = num;
        for (size_type i = 0; i != this->size(); ++i)
            this->state(i, 0).comp_num(num);
    }

    double update_log_prior (gmm_param &param) const;

    double update_log_likelihood (gmm_param &param) const
    {
        static const double log2pi = 1.8378770664093455; // log(2pi)
        double ll = -0.5 * obs_.size() * log2pi;
        param.update_log_lambda();
        for (std::size_t k = 0; k != obs_.size(); ++k) {
            double lli = 0;
            for (std::size_t i = 0; i != param.comp_num(); ++i) {
                double resid = obs_[k] - param.mu(i);
                lli += param.weight(i) * std::exp(
                        0.5 * param.log_lambda(i) -
                        0.5 * param.lambda(i) * resid * resid);
            }
            ll += std::log(lli);
        }

        return param.log_likelihood() = ll;
    }

    void read_data (const char *filename);

    private :

    NormalizingConstant nc_;
    std::size_t comp_num_;
    double alpha_, alpha_inc_;
    double mu0_, sd0_, shape0_, scale0_;
    double mu_sd_, lambda_sd_, weight_sd_;
    std::vector<double> obs_;
};
\end{Code}
The variable \code{alpha_inc_} is $\Delta\alpha(t/T_r) = \alpha(t/T_r) -
\alpha((t-1)/T_r)$, which will be used when we update the weights. The
variable \code{nc_} of type \code{NormalizingConstant} will be updated when
the weights are changed by \code{gmm_move_smc} and it will compute the
standard normalizing constant estimate
$\hat\lambda_{\text{\textsc{ds}}}^{T_r,N}$. The variables \code{mu0_} and
\code{sd0_} are the prior parameters of the means $\mu_{1:r}$. The variables
\code{shape0_} and \code{scale0_} are the prior parameters of the precisions
$\lambda_{1:r}$. The variables \code{mu_sd_}, \code{lambda_sd_}, and
\code{weight_sd_} are the proposal scales of the three random walks,
respectively. The data access member functions of these variables are omitted
in the above source code snippet.

In the \code{update_log_likelihood} member function, the calculation is a
straightforward translation of the mathematical formulation. The
\code{gmm_param::update_log_lambda} member function is used before the loop,
which simply calculates $\log\lambda_j$ for $j = 1,\dots,r$, and stores their
values. The purpose is to avoid repeated computation of these quantities
inside the loop. When the function returns, it uses the mutable version of the
\code{gmm_param::log_likelihood} member function to update the log-likelihood
stored in the \code{param} object. This is the reason that the function is
named with a \code{update} prefix. As we will see later, whenever the
parameter values are updated, it will be followed by a call to
\code{update_log_likelihood} and \code{update_log_prior}, which is implemented
in a similar fashion. Therefore the value we get by calling
\code{gmm_param::log_likelihood} will always be ``up-to-date'' while no
repeated computation is involved. Surely there are other and possibly better
design choices. However, for this simple example, this design serves our
purpose well.

The initialization is implemented using the \code{gmm_init} class,
\begin{Code}
class gmm_init : public BASE_INIT<gmm_state, gmm_init>
{
    public :

    std::size_t initialize_state (vsmc::SingleParticle<gmm_state> sp)
    {
        const gmm_state &state = sp.particle().value();
        gmm_param &param = sp.state(0);

        vsmc::cxx11::normal_distribution<> rmu(
                state.mu0(), state.sd0());
        vsmc::cxx11::gamma_distribution<> rlambda(
                state.shape0(), state.scale0());
        vsmc::cxx11::gamma_distribution<> rweight(1, 1);

        double sum = 0;
        for (std::size_t i = 0; i != param.comp_num(); ++i) {
            param.mu(i)     = rmu(sp.rng());
            param.lambda(i) = rlambda(sp.rng());
            param.weight(i) = rweight(sp.rng());
            sum += param.weight(i);
        }
        for (std::size_t i = 0; i != param.comp_num(); ++i)
            param.weight(i) /= sum;

        state.update_log_prior(param);
        state.update_log_likelihood(param);

        return 1;
    }

    void initialize_param (vsmc::Particle<gmm_state> &particle,
                           void *filename)
    {
        if (filename)
            particle.value().read_data(static_cast<const char *>(filename));
        particle.value().alpha(0);
        particle.set_equal_weight();
        particle.value().nc().initialize();
    }
};
\end{Code}
The \code{initialize_param} member function is called before the
\code{pre_processor}, which is absent in this case and have the default
implementation which does nothing. And it processes the optional parameter of
\code{Sampler::initialize}, the file name of the data. The
\code{initialize_state} member function initialize the state values according
to the prior and update the log-prior and log-likelihood.

After initialization, at each iteration, \code{gmm_move_smc} class will
implement the updating of weights as well as selecting of the proposal scales
and the distribution parameter. For example, when using the linear annealing
scheme, we can implement a \code{gmm_alpha_linear} class as the following,
\begin{Code}
class gmm_alpha_linear
{
    public :

    gmm_alpha_linear (const std::size_t iter_num) : iter_num_(iter_num) {}

    void operator() (std::size_t iter, Particle<gmm_state> &particle)
    {particle.value().alpha(static_cast<double>(iter) / iter_num_);}

    private :

    std::size_t iter_num_;
};
\end{Code}
It accepts the total number of iterations $T_r$ as an argument to its
constructor. And it implements an \code{operator()} that update the
distribution parameter $\alpha(t/T_r)$. The prior annealing scheme can be
implemented similarly. For simplicity and demonstration purpose, we only allow
\code{gmm_move_smc} to be configured with different annealing schemes, and
hard code the proposal scales. An industry strength design may make this class
a template with annealing scheme and proposal scales as policy template
parameters.
\begin{Code}
class gmm_move_smc
{
    public :

    typedef cxx11::function<void (std::size_t, Particle<gmm_state> &)>
        alpha_setter_type;

    gmm_move_smc (const alpha_setter_type &alpha_setter) :
        alpha_setter_(alpha_setter) {}

    std::size_t operator() (std::size_t iter, Particle<gmm_state> &particle)
    {
        alpha_setter_(iter, particle);

        double alpha = particle.value().alpha();
        alpha = alpha < 0.02 ? 0.02 : alpha;
        particle.value().mu_sd(0.15 / alpha);
        particle.value().lambda_sd((1 + std::sqrt(1 / alpha)) * 0.15);
        particle.value().weight_sd((1 + std::sqrt(1 / alpha)) * 0.2);

        incw_.resize(particle.size());
        weight_.resize(particle.size());
        particle.read_weight(weight_.begin());
        double coeff = particle.value().alpha_inc();
        for (vsmc::Particle<gmm_state>::size_type i = 0;
                i != particle.size(); ++i) {
            incw_[i] =
                coeff * particle.value().state(i, 0).log_likelihood();
        }
        particle.value().nc().add_log_weight(&incw_[0],
                particle.weight_set());
        particle.weight_set().add_log_weight(&incw_[0]);

        return 0;
    }

    private :

    alpha_setter_type alpha_setter_;
    std::vector<double> incw_;
    std::vector<double> weight_;
};
\end{Code}
Note that, \code{cxx11::function} is an alias to either \code{std::function}
or \code{boost::function}, depending on the value of the configuration macro
\code{VSMC_HAS_CXX11LIB_FUNCTIONAL}. Objects of this class type can be added
to a sampler as a move. The \code{operator()} satisfies the interface
requirement of the core module. First it uses \code{alpha_setter_} to set the
distribution parameter $\alpha(t/T_r)$. Second, it sets the proposal scales
for the three Metropolis random walks according to the current value of
$\alpha$. Then it computes the \emph{unnormalized} incremental weights. The
\code{NormalizingConstnat} class has member function \code{add_log_weight},
which is not unlike the one with the same name in \code{WeightSet}. It accepts
the logarithm of the incremental weights and a \code{WeightSet} object. The
standard normalizing constant estimates will be computed using these values.
The last, we also modify the \code{WeightSet} type object itself by adding the
logarithm of the incremental weights.

At each iteration, ramdom walks are also perfoemd. The implementations of the
random walks are straightforward. Below is the implementation of the random
walk on the mean parameters. The random walks on the other parameters are
similar.
\begin{Code}
class gmm_move_mu : public BASE_MOVE<gmm_state, gmm_move_mu>
{
    public :

    std::size_t move_state (std::size_t iter, SingleParticle<gmm_state> sp)
    {
        const gmm_state &state = sp.particle().value();
        gmm_param &param = sp.state(0);

        cxx11::normal_distribution<> rmu(0, state.mu_sd());
        cxx11::uniform_real_distribution<> runif(0, 1);

        double p =
            param.log_prior() + state.alpha() * param.log_likelihood();
        param.save_old();
        for (std::size_t i = 0; i != param.comp_num(); ++i)
            param.mu(i) += rmu(sp.rng());
        p = state.update_log_prior(param) +
            state.alpha() * state.update_log_likelihood(param) - p;
        double u = std::log(runif(sp.rng()));

        return param.mh_reject_mu(p, u);
    }
};
\end{Code}
First we save the logarithm of the value of target density computed using the
old values in \code{p}, the acceptance probability. And then we call
\code{gmm_param::save_old} to save the old values. Next we update each
parameter with a proposed Normal random variates and compute the new log-prior
and the log-likelihood as well as the new value of the target density. Then we
reject it according to the Metropolis algorithm, as implemented in
\code{gmm_param}, which manages both the current states as well as the backup.

Last we need to monitor certain quantities for interference purpose. Recall
that, in the \code{main} function we used
\code{sampler.path_sampling(gmm_path())} to set the monitoring of path
sampling integrands. The \code{path_sampling} member function requires a
callable objects with the following signature,
\begin{Code}
double path_eval (std::size_t iter, const Particle<T> &, double *res);
\end{Code}
The input parameter \code{iter} is the iteration number, the value of $t$ in
Equation~\ref{eq:path_est}. The return value shall be the value of $\alpha_t$.
The output parameter \code{res} shall store the array of values of
\begin{equation*}
  \rnd{\log q_{\alpha}(X_t^{(i)})}{\alpha}\Bigm|_{\alpha = \alpha_t}.
\end{equation*}
Our implementation of \code{gmm_path} is a sub-class of an \smp module base
class, which provides an \code{operator()} that satisfies the above interface
requirement. Its usage is similar to the \code{MoveSEQ} template introduced in
Section~\ref{sub:SMP module}.

The path sampling integrands under this geometry annealing scheme are simply
the log-likelihood. Therefore the implementation of \code{gmm_path} class is
rather simple,
\begin{Code}
class gmm_path : public BASE_PATH<gmm_state, gmm_path>
{
    public :

    double path_state (std::size_t, ConstSingleParticle<gmm_state> sp)
    {return sp.state(0).log_likelihood();}

    double path_grid (std::size_t, const Particle<gmm_state> &particle)
    {return particle.value().alpha();}
};
\end{Code}

\subsubsection{Results}

After compiling and running the algorithm, the results were consistent with
those reported in \citet{DelMoral:2006hc}. For a more in depth analyze of the
methodologies, extensions and the results see \citet{Zhou2013mc}.

\subsubsection[Extending the implementation using MPI]{Extending the
  implementation using \mpi}

\vsmc's \mpi module assumes that identical samplers are constructed on each
node, with possible different number of particles to accommodate the
difference in capacities among nodes. To extend the above \smp implementation
for use with \mpi, first at the beginning of the \code{main} function, we add
the following,
\begin{Code}
MPIEnvironment env(argc, argv);
\end{Code}
to initialize the \mpi environment. When the object \code{env} is destroyed at
the exit of the \code{main} function, the \mpi environment is finalized.
Second, we need to replace base value collection class template with
\code{StateMPI}. So now \code{gmm_state} is declared as the following,
\begin{Code}
class gmm_state :
    public StateMPI<BASE_STATE<StateMatrix<RowMajor, 1, gmm_param> > >;
\end{Code}
The implementation is exactly the same as before. Third, the \code{gmm_param}
class now needs to be transferable using \mpi. Unlike the \smp situations, a
simple copy constructor is not enough. \vsmc uses \boost{} \mpi library, and
thus one only needs to write a \code{serialize} member function for
\code{gmm_param} such that the data can be serialized into bytes. See
documents of \boost{} \mpi and serialization libraries for details. In
summary, the following member function accept an \code{Archive} object as
input, and it can perform a store or a load operation based on the
\code{Archive} type. In a load operation, the \code{Archive} object is like an
input stream and in a store operation, it is like an output stream.
\begin{Code}
template <typename Archive>
void serialize (Archive &ar, const unsigned)
{
    int num = comp_num_;
    ar & num;
    comp_num(num);

    ar & log_prior_;
    ar & log_likelihood_;
    for (std::size_t i = 0; i != comp_num_; ++i) {
        ar & mu_[i];
        ar & lambda_[i];
        ar & weight_[i];
        ar & log_lambda_[i];
    }
}
\end{Code}
Fourth, after user input of the sampler parameters, we need to sync them with
all nodes. For example, for the \code{ParticleNum} parameter,
\begin{Code}
boost::mpi::communicator World;
boost::mpi::broadcast(World, ParticleNum, 0);
\end{Code}
Last, any importance sampling estimates that are computed on each node, need
to be combined into final results. For example, the path sampling results are
now obtained through adding the results from each node together,
\begin{Code}
double ps_sum = 0;
boost::mpi::reduce(World, ps, ps_sum, std::plus<double>(), 0);
ps = ps_sum;
\end{Code}
For the standard normalizing constant ratio estimator, we will replace
\code{NormalizingConstant} with \code{NormalizingConstantMPI}, which will
perform such and other tasks.

After these few lines of change, the sampler is now parallelized using \mpi
and can be deployed to clusters and other distributed memory architecture. On
each node, the selected \smp parallelization is used to perform
multi-threading parallelization locally. \vsmc's \mpi module will take care of
normalizing weights and other tasks.

\subsubsection{Parallelization performance}

One of the main motivation behind the creation of \vsmc is to ease the
parallelization with different programming models. The same implementation can
be used to built different samplers based on what kind of parallel programming
model is supported on the users' platforms. In this section we compare the
performance of various \smp parallel programming models and \opencl
parallelization.

We consider five different implementations supported by \icpc 2013:
sequential, \tbb, \cilk, \openmp and \cppoo{} \code{<thread>}. The samplers
are compiled with
\begin{Code}
CXX=icpc -std=c++11 -gcc-name=gcc-4.7 -gxx-name=g++-4.7
CXXFLAGS=-O3 -xHost -fp-model precise  \
         -DVSMC_HAS_CXX11LIB_FUNCTIONAL=1  \
         -DVSMC_HAS_CXX11LIB_RANDOM=1
\end{Code}
on a Ubuntu 12.10 workstation with an Xeon W3550 (3.06GHz, 4 cores, 8 hardware
threads through hyper-threading) \cpu. A four components model and $100$
iterations with a prior annealing scheme is used for all implementations. A
range of numbers of particles are tested, from $2^3$ to $2^{17}$.

\begin{figure}
  \centering
  \includegraphics[width=\linewidth]{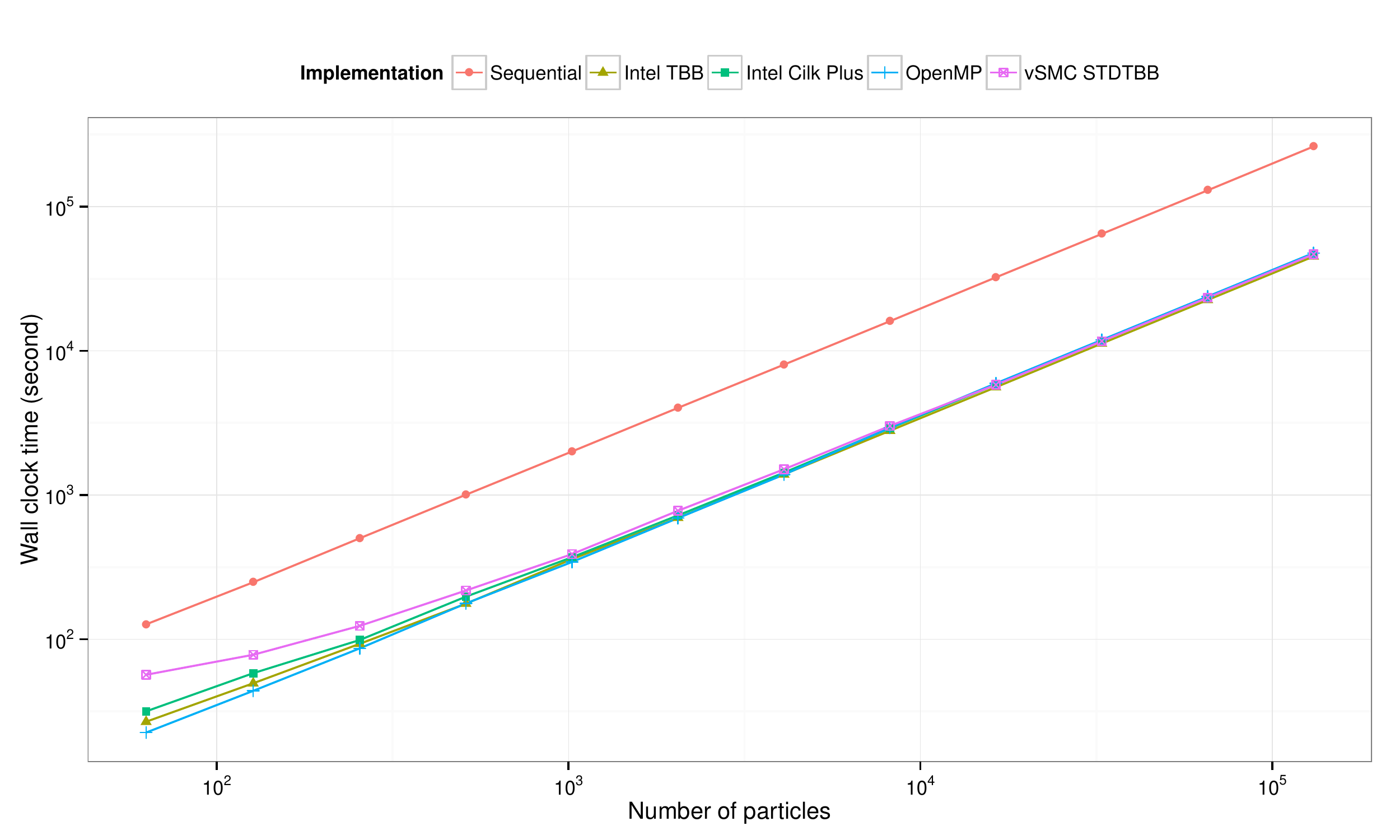}
  \includegraphics[width=\linewidth]{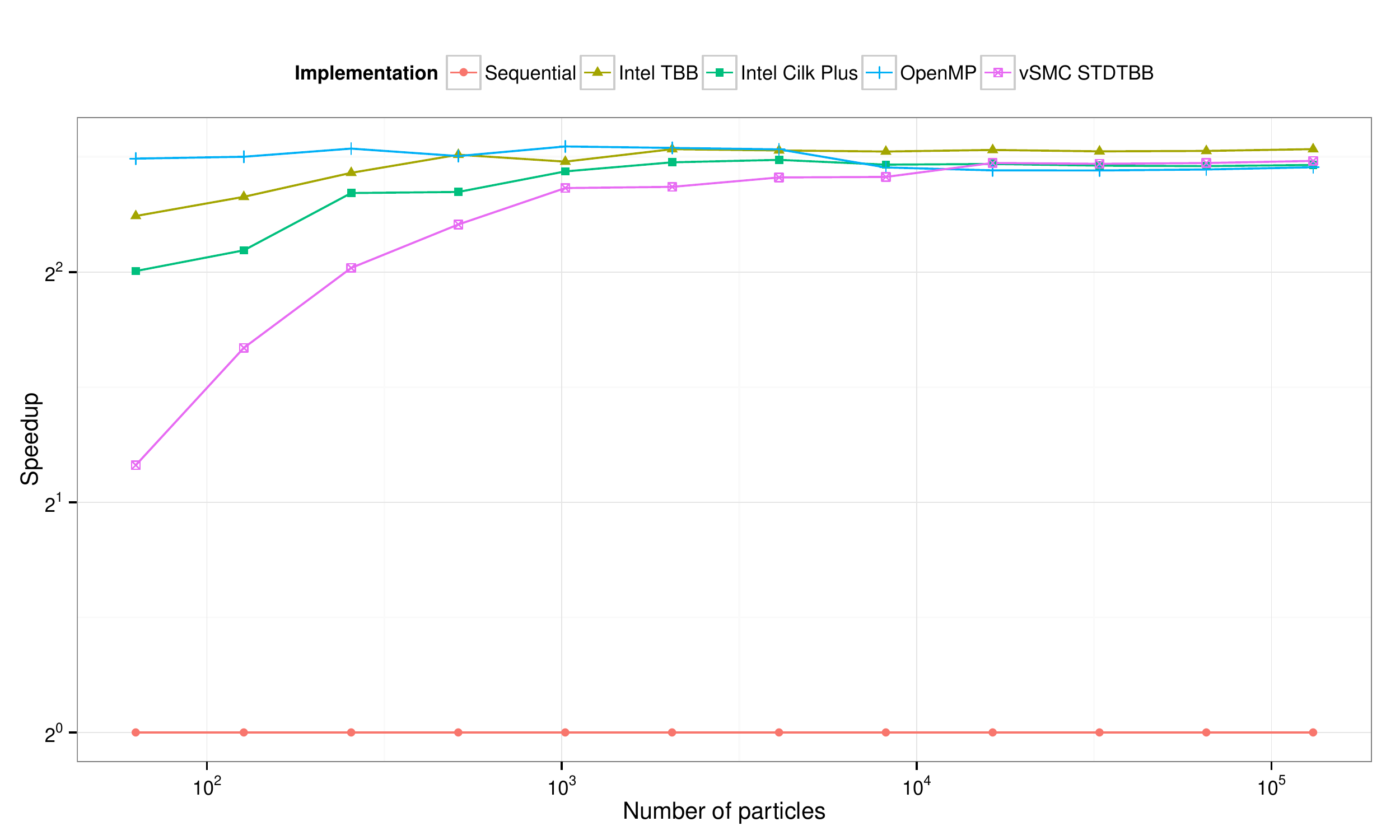}
  \caption{Performance of \cpp implementations of Bayesian modeling for
    Gaussian mixture model (Linux; Xeon W3550, 3.06GHz, 4 cores, 8 threads).}
  \label{fig:bench-smp-perf}
\end{figure}

For different number of particles, the wall clock time and speedup are shown
in Figure~\ref{fig:bench-smp-perf}. For $10^4$ or more particles, the
differences are minimal among all the programming models. They all have
roughly 550\% speedup. With smaller number of particles, \vsmc's \cppoo
parallelization is less efficient than other industry strength programming
models. However, with $1000$ or more particles, which is less than typical
applications, the difference is not very significant.

\opencl implementations are also compared on the same workstation, which also
has an NVIDIA Quadro 2000 graphic card. \opencl programs can be compiled to
run on both \cpu{}s and \gpu{}s. For \cpu implementation, there are \ciocl and
\caocl platforms. We use the \tbb implementation as a baseline for comparison.
The same \opencl implementation are used for all the \cpu and \gpu runtimes.
Therefore they are not particularly optimized for any of them. For the \gpu
implementation, in addition to double precision, we also tested a single
precision configuration.  Unlike modern \cpu{}s, which have the same
performance for double and single precision floating point operations (unless
\simd instructions are used, which can have at most a speedup by a factor of
2), \gpu{}s penalize double precision performance heavily.

\begin{figure}
  \centering
  \includegraphics[width=\linewidth]{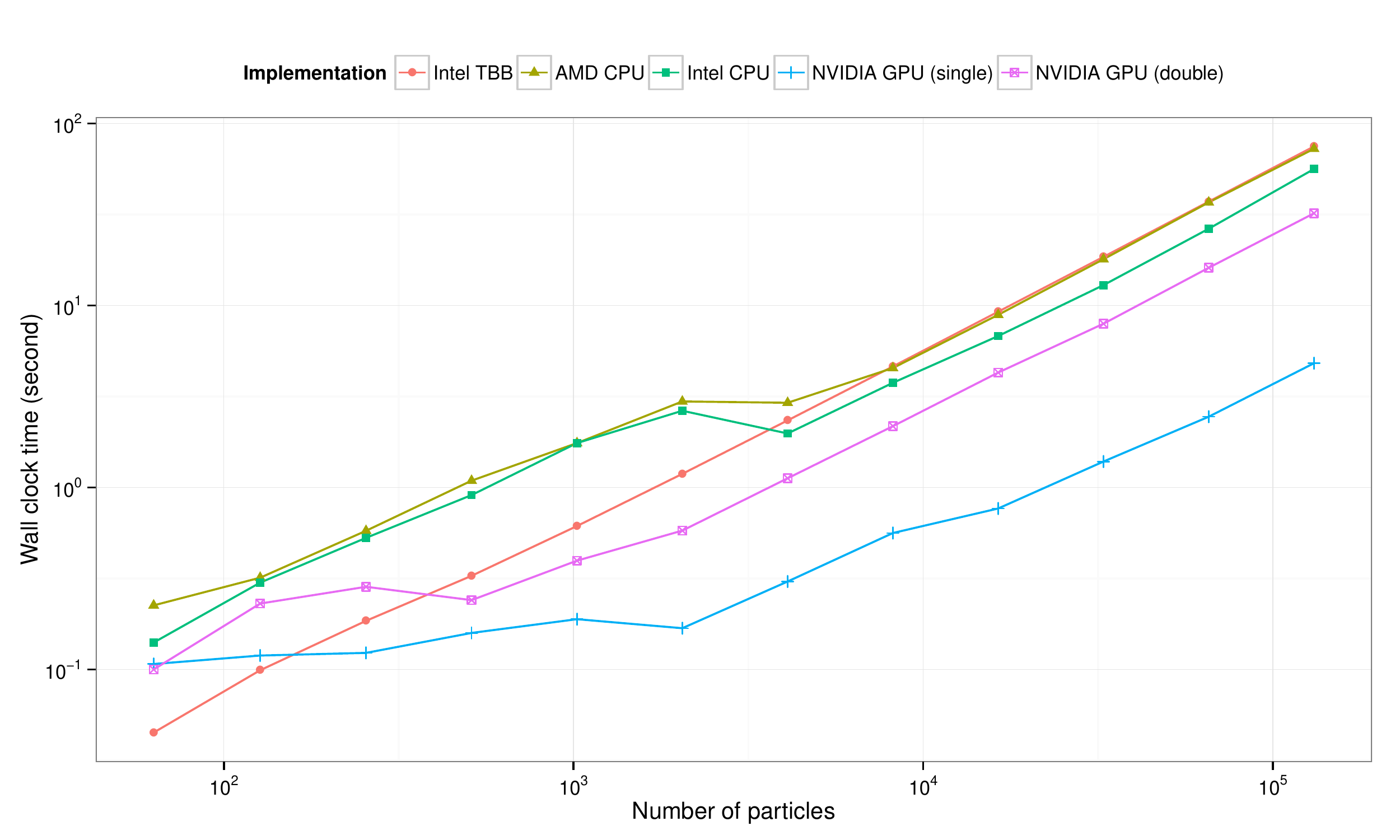}
  \includegraphics[width=\linewidth]{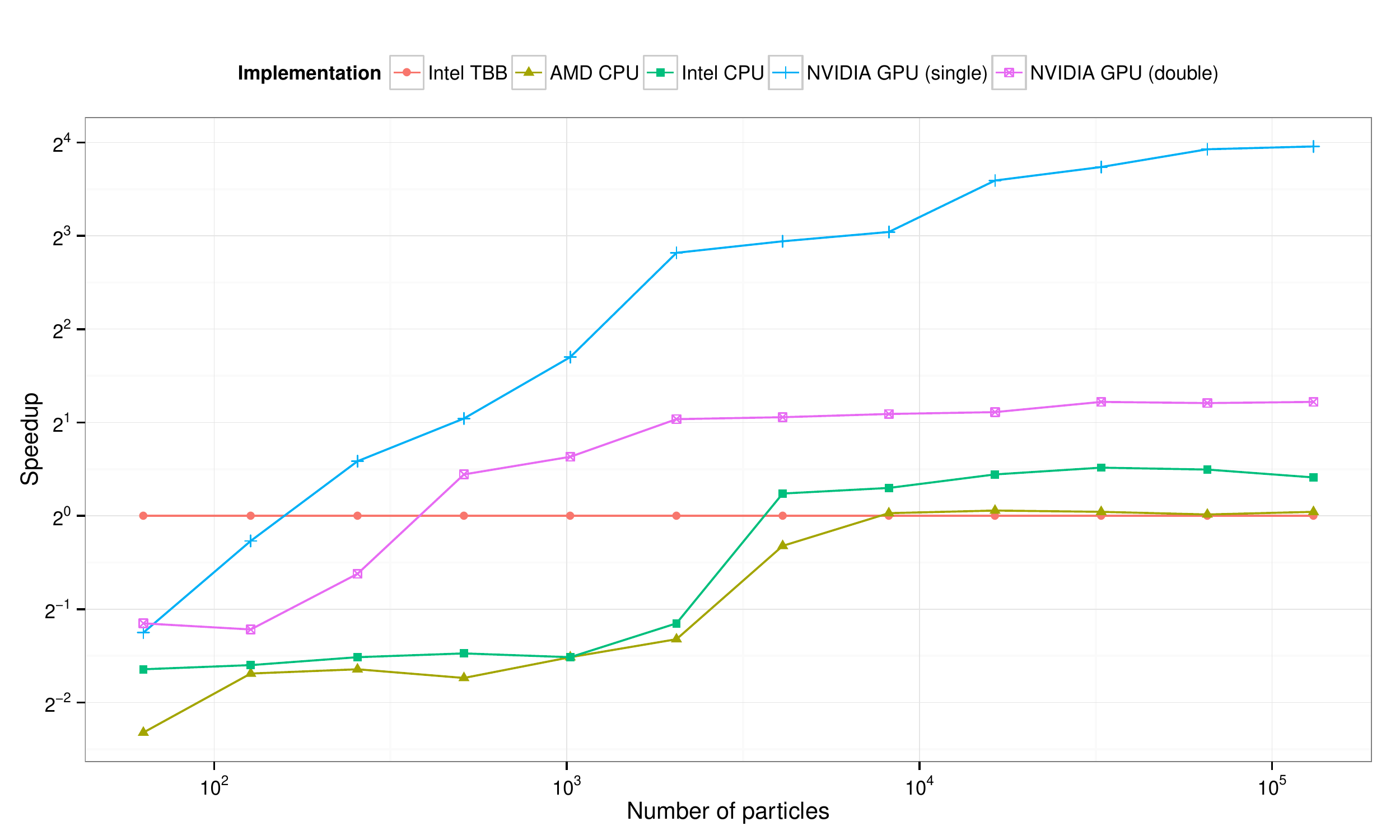}
  \caption{Performance of \opencl implementations of Bayesian modeling for
    Gaussian mixture model (Linux; Xeon W3550 \gpu, 3.06GHz, 4 cores, 8
    threads; NVIDIA Quadro 2000).}
  \label{fig:bench-ocl-perf}
\end{figure}

For different number of particles, the wall clock time and speed up are
plotted in Figure~\ref{fig:bench-ocl-perf}. With smaller number of particles,
the \opencl implementations have a high overhead when compared to the \tbb
implementation. With a large number of particles, \aocl has a similar
performance as the \tbb implementation. \iocl is about 40\% faster than the
\tbb implementation. This is due to more efficient vectorization and compiler
optimizations. The double precision performance of the NVIDIA \gpu has a 220\%
speedup and the single precision performance has near 1600\% speedup. As a
rough reference for the expected performance gain, the \cpu has a theoretical
peak performance of 24.48 GFLOPS. The \gpu has a theoretical peak performance
of 60 GFLOPS in double precision and 480 GFLOPS in single precision. This
represents 245\% and 1960\% speedup compared to the \cpu, respectively.

It is widely believed that \opencl programming is tedious and hard. However,
\vsmc provides facilities to manage \opencl platforms and devices as well as
common operations. Limited by the scope of this paper, the \opencl
implementation (distributed with the \vsmc source) is not documented in this
paper. Overall the \opencl implementation has about 800 lines including both
host and device code. It is not an enormous increase in effort when compared
to the 500 lines \smp implementation. Less than doubling the code base but
gaining more than 15 times performance speedup, we consider the programming
effort is relatively small.

\section{Discussion}
\label{sec:Discussion}

This paper introduced a \cpp template library intended for implementing
general \smc algorithms and constructing parallel samplers with different
programming models. While it is possible to implement many realistic
application with the presented framework, some technical proficiency is still
required to implement some problem specific part of the algorithms. Some basic
knowledge of \cpp in general and how to use a template library are also
required.

It is shown that with the presented framework it is possible to implement
parallelized, scalable \smc samplers in an efficient and reusable way. The
performance of some common parallel programming models are compared using an
example.

Some future work may worth the effort to ease the implementation of \smc
algorithms further. However, there is a balance between performance,
flexibility and the ease of use. \vsmc aims to be developer-friendly and to
provide users as much control as possible for all performance related aspects.
For a \bugs-like interface, users may be interested in other software such as
\cbiips. In addition \clibbi provides a user friendly and high performance
alternative with a focus on state-space models. Compared with these recent
developments, \vsmc is less accessible to those with little or no knowledge of
\cpp. However, for researchers with expertise in \cpp and template generic
programming in particular, \vsmc provides a framework within which
potential superior performance can be obtained and greater flexibility and
extensibility are possible.

\bibliography{paper}
\end{document}